\documentclass[10pt,journal,letterpaper,twocolumn,twoside]{IEEEtran} 

\usepackage[dvips]{epsfig}
\usepackage[dvips]{graphicx}
\usepackage{amsmath,amsfonts,bm,amssymb,psfrag,ifthen,color,subfigure}
\usepackage{algpseudocode}
\usepackage{algorithm}
\usepackage{algorithmicx}
\usepackage{setspace}
\usepackage{cite}
\usepackage{url}
\usepackage{longtable}
\usepackage{stfloats}  
\usepackage{graphics,booktabs,color,subfigure}
\usepackage{float}
\usepackage{diagbox}
\usepackage[justification=centering]{caption}
\usepackage{multirow}
\usepackage{array}

\begin{document}
 \title{Optimal Probabilistic Constellation Shaping  for  Covert Communications}

\author{Shuai Ma, Yunqi Zhang, Haihong Sheng, Hang Li, Jia Shi, Long Yang, Youlong Wu,    Naofal Al-Dhahir,~ and Shiyin Li

\thanks{Shuai Ma,Yunqi Zhang, Haihong Sheng and Shiyin Li   are with the School of Information and Control   Engineering, China
University of Mining and Technology, Xuzhou 221116,
China (e-mail: mashuai001@cumt.edu.cn).}

}

\maketitle
 \begin{abstract}
In this paper, we investigate the optimal probabilistic constellation shaping design for
   covert communication  systems  from a practical view.
   Different from conventional  covert communications with equiprobable constellations modulation,
      we propose   non-equiprobable constellations modulation schemes to further enhance the covert rate.
   Specifically, we   derive covert
rate expressions  for practical discrete constellation inputs for the
first time. Then, we study the covert rate maximization problem by jointly
optimizing the constellation distribution and power allocation. In particular,  an approximate gradient descent method
is proposed for obtaining the optimal probabilistic constellation shaping.
To strike a balance between the computational complexity and
the transmission performance, we further develop a framework
that maximizes a lower bound on the achievable rate where
the optimal probabilistic constellation shaping problem  can be solved efficiently
using the Frank-Wolfe method. Extensive numerical results show that
the optimized probabilistic constellation shaping  strategies  provide significant gains in the achievable covert rate over the state-of-the-art schemes.

\end{abstract}

\IEEEpeerreviewmaketitle

\section{Introduction}

Radio frequency (RF) based wireless communication is inevitably susceptible to eavesdropping due to the broadcast nature of the electromagnetic waves.
With the  ever-growing      Internet of Things (IoT) applications, the security issue has become more and more crucial in the future sixth-generation
(6G) wireless networks \cite{Nguyen21Security}. For example,  in large enterprise buildings, hospitals, factories,  communication may be very sensitive to a hostile adversary.
  The conventional cryptography
approaches \cite{Chen17Survey} usually focus on  protecting  the transmission content or increasing message decoding complexity, while the  physical-layer security \cite{Barros11Physical, Wang16Physical} approaches exploit the intrinsic wireless fading channels  properties
  to minimize
the information leakage to the eavesdroppers.
In fact, a higher level of security is to hide the existence of the communication, which not only can be applied in all aforementioned application scenarios {\cite{Du2021Optimal,Du2022Reconfigurable,Du2022performance,Xie2022sec}}, but also meets more critical demands from military or security agencies.
To address this high level security, covert communications \cite{Yan2019Gaussian,Bloch16}, which shields the  existence of message transmissions against the detection of a warden,
are emerging  as a cutting-edge wireless communication security technique, and have recently attracted significant research attention. {{Note that, covert communication aims to hide the communication behavior from the eavesdropper, while physical layer security tries to reduce the interception information of the eavesdropper.
}}


The basic idea of covert communications is as follows. The legitimate transmitter (Alice) transmits messages to the paired receiver (Bob), while   guaranteeing a low detection probability for a Warden (Willie).
  Although such idea has been realized by  spread-spectrum techniques \cite{Simon94}
for several decades, the information-theoretic limits of covert communications, which   is also referred to as low  probability of detection (LPD) communications in some literature,  were only recently derived \cite{Bash13,Bloch16,Wornell16,Wu17}. In particular, the authors in \cite{Bash13} firstly demonstrated that  in  additive white Gaussian noise
(AWGN) channels,  Alice can reliably
send at most $\mathcal{O}\left(\sqrt n\right)$ bits to Bob in $n$ channel usages under the covert requirement. Such result is also called the square root law (SRL).
Subsequent works have extended this result
   to various channel models such as    binary symmetric channels    \cite{Che2013Reliable}, broadcast communications \cite{Arumugam_TIFS_2019},   multiple access channels \cite{Arumugam_TIT_2019}, and interference channels\cite{Cho_TIFS_2020}.

Although the SRL indicates that  the asymptotic achievable rate under the covert requirement approaches zero,
     many researchers have shown that  the SRL limit  can be beaten by exploiting additional techniques in the considered covert communication scenario. These methods include: taking  advantage of the ignorance of the transmission time at  Willie \cite{Bash2016Covert}; applying an intelligent reflecting surface \cite{Wang_TOM_2021,Chen_WCL_2021,Si_TCOM_2021}; exploring the molecular absorption   or scattering feature of the Terahertz
spectrum \cite{Gao_TWC_2020,Liu_IoT_2020}; cooperative jamming \cite{Zheng_TWC_2021} or uninformed jamming \cite{Sobers2017Covert,Li_TWC_2020,Shmuel_TCOM_2021,Li_TWC_2021}; jointly optimizing the beam
training  and data transmission for millimeter-wave communication \cite{Zhang_TIFS_2021};
 exploiting the  uncertainty  noise power or channel state information (CSI)    at Willie    \cite{Goeckel16cov,Lee15ach,Yan17on,Shahzad_TIFS_2020,Cheng_TCOM_2021}; robust beamforming design \cite{Zheng2019Multi-Antenna,MA_TIFS_2021},  over a finite number of
channel uses\cite{Yan19Delay};    applying a   full-duplex transceiver \cite{Shu2019Delay,Wang_TWC_2021,SUN_TCOM_2021,Zheng_TCOM_2020}; intermediate relay \cite{Hu2019Covert,Sheikholeslami2018Multi-Hop} or exploring   unmanned aerial
vehicle (UAV) as mobile relay \cite{WangHM_TWC_2019,Yan_ISAC_2021}.

        To be more specific,  it is shown in \cite{Bash2016Covert} that Alice can covertly  transmit ${\cal O}\left( {\min \left\{ {n,\sqrt {n\log T\left( n \right)} } \right\}} \right)$
bits to Bob.
When Willie lacks the knowledge of his noise power, Alice can reliably transmit  $\mathcal{O}\left( n\right)$
bits  \cite{Goeckel16cov,Lee15ach,Yan17on}.
With the aid of   an uninformed jammer   \cite{Sobers2017Covert},
Alice can also achieve  positive transmission rate.
 With a finite number of channel uses, a uniformly distributed power allocation scheme was proposed \cite{Yan19Delay}  to enhance the covert transmission.
 In \cite{Shu2019Delay}, the authors showed that, the effective throughput under delay constraints can be improved by adding artificial noise (AN)  at the  full-duplex  receiver.
For a one-way relay network, the authors in \cite{Hu2019Covert} studied   the performance limits of convert communications  of an energy harvesting relay.
In \cite{Sheikholeslami2018Multi-Hop},  a multiple-relay network was considered, and both the maximum throughput and the minimum
end-to-end delay routing algorithms were developed  with  multiple Willies.
 In \cite{Zheng2019Multi-Antenna}, two probabilistic metrics, called the covert outage
probability and the connectivity probability, were analyzed for  multi-antenna  covert communications
with randomly located wardens and
interferers.
 Using Kullback-Leibler divergence, i.e., $D\left( {{p_1}||{p_0}} \right)$ or $D\left( {{p_0}||{p_1}} \right)$
 to measure the covertness,  a  Gaussian input distribution
  was shown to be optimal for the covert  metric $D\left( {{p_1}||{p_0}} \right)$, and not   optimal for the covert  metric $D\left( {{p_0}||{p_1}} \right)$\cite{Yan2019Gaussian},
where ${p_0}$ and ${p_1}$ represent Willie's received  signal distributions  when
covert communications occur   and  not  occur, respectively.

The aforementioned research advances in covert communications
mainly make the assumption of a Gaussian input distribution at the transmitter side,
which can hardly be realized in practical communication
systems.
In fact, the information symbols in practical communication systems are realized in
the form of  discrete constellation   points, i.e., finite alphabet inputs, such as pulse amplitude modulation (PAM)
  and multiple-quadrature amplitude modulation (M-QAM).
In \cite{Topal_WCL_2020}, both the lower bound of  covert transmission probability and throughput maximization have been analyzed  with    discrete constellation inputs.  The discrete  constellation points are assumed to be equally likely, which  is not   optimal for practical covert communications, especially for high-order modulation schemes.
 So far, the optimal  discrete  constellation inputs of covert communication are still not well discussed in the literature.

Motivated by the above background, we     develop   information-theoretic limits of covert communications  with probabilistic constellation shaping.  First, we
derive the achievable rate expression of the system with the
discrete constellation input signals, rather than the Gaussian
inputs adopted in most of the existing works. Then, we
investigate the  performance with the optimal input distribution. Our results provide a
practical design framework for covert communication systems.
 The main contributions of this
paper are summarized as follows:
\begin{itemize}
\item Generally, the inputs of practical communications
systems follow a finite-set discrete distribution rather
than a Gaussian distribution. To evaluate performance, we derive the achievable
rate expressions for an arbitrary discrete distributed input. Comparing to the existing rate expressions with
equiprobable discrete constellation points, the derived
expressions are more general and practical. Since the derived
rate expression is not in closed-form, we further derive
both lower and  upper bounds. All these results
can be used as  performance metrics for the considered covert communication
system.
\item
Furthermore, we design optimal  discrete constellation inputs to
 maximize the  exact covert  rate under the covertness constraints,  transmit    power limitations, and the signal distribution requirements, which is a challenging problem since neither the exact covert  rate nor the covertness constraint has an analytical expression. To
efficiently solve it, we  conservatively transform the covertness constraint into its upper bound with closed-form expression.
 Then, we adopt the numerical integration method
to approximate the  covert  rate objective function and its gradient.
Afterwards,   the optimal
probability distributions of the discrete constellation are
calculated by the approximate gradient descent method, where   the step sizes are calculated
by the backtracking line search.
\item To reduce the computation complexity of the
design problem, we further adopt the derived lower
bound as the covert  rate performance metric.  To overcome  the non-convexity challenge, this problem is iteratively
solved by the proposed Frank-Wolfe method.

 \end{itemize}

 The rest of this paper is organized as follows. The
system model and the  derivation of Bob's achievable rate are presented in Section II.
The optimal  probabilistic constellation shaping design for covert  communications  is provided in Section III.
     The  probabilistic constellation shaping     design and its approximations are presented in Section IV.
        In Section V, we evaluate the proposed  probabilistic constellation shaping design using numerical results. Finally, we conclude  the paper in Section   VI.

\emph{Notations}:  The vectors and matrices are represented by
boldfaced lowercase and uppercase letters, respectively.
The notations ${\left(  \cdot  \right)^{\rm{*}}}$, ${{\mathbb E}}\left\{  \cdot  \right\}$,
$\left\|  \cdot  \right\|$, ${\rm{Tr}}\left(  \cdot  \right)$, ${\mathop{\rm Re}\nolimits} \left(  \cdot  \right)$ and  ${\mathop{\rm Im}\nolimits} \left(  \cdot  \right)$ represent the conjugate,
the expectation,  Frobenius norm,  trace, the real part and imaginary part of its argument, respectively.
And $ \odot$ is Hadamard Product,i.e.${A_{m \times n}}\left[ {{a_{ij}}} \right] \odot {B_{m \times n}}\left[ {{b_{ij}}} \right] = {C_{m \times n}}\left[ {{a_{ij}}{b_{ij}}} \right]$.
 The operator ${\bf{A}}\underline  \succ  {\bf{0}}$ means ${\bf{A}}$ is positive semidefinite.
The notation $\mathcal{CN}\left( {\mu ,{\sigma ^2}} \right)$ denotes a complex-valued circularly symmetric Gaussian distribution with   mean   $\mu$ and  variance   ${\sigma ^2}$.

\section{System model}

\begin{figure}[h]
      \centering
	\includegraphics[width=7cm]{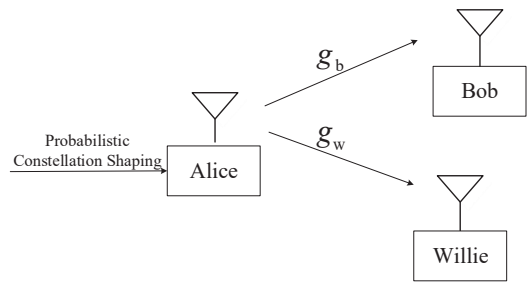}
    \caption{ {The system model of covert communication.}}
  \label{model}
\end{figure}
Consider a typical covert communication scenario as illustrated in Fig. 1, where Alice and Bob are a legitimate communication pair, and Willie is the eavesdropper. Each one of them is equipped with a single antenna.  Let ${{{g}}_{\rm{b}}} \sim {\cal CN}\left( {{{0}},\sigma_1^2} \right)$ and ${{{g}}_{\rm{w}}} \sim {\cal CN}\left( {{{0}},\sigma_2^2} \right)$ denote the  Rayleigh  flat fading channel from Alice to Bob and Willie, respectively \cite{Shahzad2017Covert}, where $\sigma_1^2$ and $\sigma_2^2$ are the variances of   ${{{g}}_{\rm{b}}}$ and ${{{g}}_{\rm{w}}}$.
Let     $x\left[ i \right]$ denote Alice's transmitted  symbol  at the $i$-th  channel  use, where $i = 1,...,N$, and $N$ is the total number of channel uses.

\subsection{ Achievable Rate of Bob}
 As it is the case in practice, $x\left[ i \right] \in \Omega $
 follows a discrete constellation distribution instead of Gaussian  distribution. Here, $\Omega$  denotes a discrete constellation set  with $K$ discrete points
 ${\left\{ {{x_k}} \right\}_{1 \le k \le K}}$, i.e.,
\begin{align} \label{Omega}
\Omega  = \left\{ {X\left| {\begin{array}{*{20}{l}}
\begin{array}{l}
\Pr \left( {X = {x_k}} \right) = {p_k} \ge {\rm{0,}}\sum\limits_{k = 1}^K {{p_k}}  = 1,\\
\sum\limits_{k = 1}^K {{p_k}} {\left| {{x_k}} \right|^2} \le {P_{\rm{A}}},{x_k} \in {\mathbb {C}},k = 1,...,K
\end{array}
\end{array}} \right.} \right\},
\end{align}
  where ${{x_k}}$  denotes the  $k$th  discrete point,    ${{p_k}}$  denotes the  corresponding probability, and ${P_\text{A}}$ denotes the average power.

 For the $i$-th  channel  use,  the received signal at Bob   ${y_{\rm{b}}}\left[ i \right]$
  is given as
\begin{align}
{y_{\rm{b}}}\left[ i \right]
 = {g_{\rm{b}}^{\rm{*}}}{x}\left[ i \right] + {z_{\rm{b}}}\left[ i \right],
\end{align}
where  ${z_{\rm{b}}}\left[ i \right] \sim \mathcal{CN}\left( {0,\sigma _{\rm{b}}^2} \right)$  denotes   the  received noise  at Bob.
Since $x\left[ i \right] \in \Omega$, then the likehood functions of  ${y_{\rm{b}}}\left[ i \right]$  is given as
\begin{align}\label{pdf}
{p}\left( {{y_{\rm{b}}}} \right) = \frac{1}{{{\pi } {{\sigma _{\rm{b}}^2}}}}\sum\limits_{k = 1}^K {{p_k}\exp \left( { - \frac{{{{\left| {{y_{\rm{b}}} - {g_{\rm{b}}^{\rm{*}}}{x_k}} \right|}^2}}}{{\sigma _{\rm{b}}^2}}} \right)}.
\end{align}

Therefore, given the discrete constellation,  the achievable rate of Bob ${R_{\rm{b}}}$    is given  by
\begin{subequations}\label{MI}
\begin{align}
&{R_{\rm{b}}}=I\left( {{y_{\rm{b}}}\left[ i \right];{x}\left[ i \right]} \right) \\
&= h\left( {{y_{\rm{b}}}\left[ i \right]} \right) - h\left( {{z_{\rm{b}}}\left[ i \right]} \right) \\
&= - \int_{ - \infty }^\infty  {{p}\left( {{y_{\rm{b}}}} \right)} {\log _2}{p}\left( {{y_{\rm{b}}}} \right)d{y_{\rm{b}}} - {\log _2}\pi e\sigma _b^2\\
&=   - \sum\limits_{k = 1}^K {{p_k}} {\mathbb{E}_{{z_{\rm{b}}}}}\left\{ {{{\log }_2}\sum\limits_{j = 1}^K {{p_j}} \exp \left( { - \frac{{{{\left| {{g_{\rm{b}}^{\rm{*}}}\left( {{x_k} - {x_j}} \right) + {z_{\rm{b}}}} \right|}^2}}}{{\sigma _{\rm{b}}^2}}} \right)} \right\}\nonumber\\
&\quad - \frac{1}{{\ln 2}},
\end{align}
\end{subequations}
{where $h\left( X \right) =  - \int {f\left( x \right)} \log f\left( x \right)dx$ denotes differential entropy, and
 ${f\left( x \right)}$ represents the probability density function (PDF).}

Based on the achievable rate expression {in \eqref{MI}}, we will further investigate the optimal   probability of discrete constellations for covert communications.
\subsection{Hypothesis Testing}

According to the received signals,
Willie attempts to decide whether Alice  covertly transmits information
to Bob or not by performing an optimal statistical hypothesis
test (such as the Neyman-Pearson test).
Specifically, Willies needs to distinguish between two hypotheses: 1) the null
hypothesis  ${{\cal{H}}_0}$  indicating  no transmission; 2) the hypothesis ${{\cal{H}}_1}$ indicating the   transmission.  Let $y_{\rm{w}}\left[ i \right]$ denote  the  received  signal  at  Willie  in the $i$-th  channel  use.
 Under the two hypotheses, the signal  received at Willie is given as
\begin{subequations}\label{hypothesis}
\begin{align}
&\quad  {{\cal{H}}_0}:y_{\rm{w}}\left[ i \right] = {z_{\rm{w}}}\left[ i \right],\\
&\quad  {{\cal{H}}_1}:y_{\rm{w}}\left[ i \right] = g_{\rm{w}}^{\rm{*}}x\left[ i \right] + {z_{\rm{w}}}\left[ i \right],
\end{align}
\end{subequations}
where
${z_{\rm{w}}}\left[ i \right]\sim \mathcal{CN}\left( {0,\sigma _{\rm{w}}^2} \right)$  denotes   the  received noise  at   Willie.
Let ${{\cal{D}}_1}$ and ${{\cal{D}}_0}$,  respectively, denote
the binary decisions of Willie.
Thus, the total detection  error probability of  Willie is defined as \cite{Yan2019Gaussian,T.M.Cover02,Lehmann_2005_Testing}
\begin{align}
\xi  = \Pr \left( {{{\cal{D}}_1}|{{\cal{H}}_0}} \right) + \Pr \left( {{{\cal{D}}_0}|{{\cal{H}}_1}} \right).
\end{align}

Note that,  $\Pr \left( {{{\cal{D}}_1}|{{\cal{H}}_0}} \right)$
  denotes the  false alarm  probability  that  Willie believes ${{\cal{H}}_1}$ when Alice   does not  transmit,
  and $\Pr \left( {{{\cal{D}}_0}|{{\cal{H}}_1}} \right)$ denotes the
   missed detection probability    that  Willie decides ${{\cal{H}}_0}$ when  Alice transmits.
Moreover, let
${p_{y,0}} = f\left( {{y_{\rm{w}}}\left| {{{\cal{H}}_0}} \right.} \right)$ and ${p_{y,1}}= f\left( {{y_{\rm{w}}}\left| {{{\cal{H}}_1}} \right.} \right)$  denote the likehood  functions  of $y_{\rm{w}}$   under ${{\cal{H}}_0}$  and  ${{\cal{H}}_1}$, respectively.
According to  \eqref{hypothesis},  we have
\begin{subequations}\label{likehood_function}
\begin{align}
&{p_{y,0}}  = \frac{1}{{ {\pi } {{\sigma _{\rm{w}}^2}}}}\exp \left( { - \frac{\left|{{y_{\rm{w}}}}\right|^2}{{\sigma _{\rm{w}}^2}}} \right),\\
&{p_{y,1}} = \frac{1}{{ {\pi } {{\sigma _{\rm{w}}^2}}}}\sum\limits_{k = 1}^K {{p_k}\exp \left( { - \frac{{{{\left| {y_{\rm{w}} - {g_{\rm{w}}^{\rm{*}}}{x_k}} \right|}^2}}}{{\sigma _{\rm{w}}^2}}} \right)}.
\end{align}
\end{subequations}
Let  ${V_T}\left( {{{p_{y,0}}\left\| {{p_{y,1}}} \right.}} \right) = \frac{1}{2}{\left\| {{p_{y,0}} - {p_{y,1}}} \right\|_1}$ denote
the total variation distance between ${p_{y,0}}$  and ${p_{y,1}}$.
According to Theorem 13.1.1 in  \cite{Lehmann_2005_Testing},  the  optimal   detection error probability  of  Willie is given as
\begin{align}
{\xi ^ {\rm{opt}} } = 1 - {V_T}\left( {{{p_{y,0}}\left\| {{p_{y,1}}} \right.}} \right)  = 1 - \frac{1}{2}{\left\| {{p_{y,0}} - {p_{y,1}}} \right\|_1}.
\end{align}

However, in general, ${V_T}\left( {{{p_{y,0}}\left\| {{p_{y,1}}} \right.}} \right)$  is   difficult to analyze. To address this issue,
we apply  Pinsker's inequality \cite{T.M.Cover02} to obtain an upper bound
\begin{align}
 &{V_T}\left( {{{p_{y,0}}\left\| {{p_{y,1}}} \right.}} \right) \le \sqrt {\frac{1}{2}D\left( {{p_{y,0}}\left\| {{p_{y,1}}} \right.} \right)},  \\
 &{V_T}\left( {{{p_{y,0}}\left\| {{p_{y,1}}} \right.}} \right) \le \sqrt {\frac{1}{2}D\left( {{p_{y,1}}\left\| {{p_{y,0}}} \right.} \right)},
\end{align}
where $D\left( {{p_{y,0}}\left\| {{p_{y,1}}} \right.} \right) = \int_y {{p_{y,0}}{{\log }_2}\frac{{{p_{y,0}}}}{{{p_{y,1}}}}} d{y_{\rm{w}}}$ denotes
 the  Kullback-Leibler (KL)  divergence from ${p_{y,0}}$ to  ${p_{y,1}}$, and $ D\left( {{p_{y,1}}\left\| {{p_{y,0}}} \right.} \right)  = \int_y {{p_{y,1}}{\log _2} \frac{{p_{y,1}}}{{p_{y,0}}}} d{y_{\rm{w}}}$
  denotes the KL divergence  from ${p_{y,1}}$ to  ${p_{y,0}}$.

Based on the  likehood functions of $y_{\rm{w}}$  in \eqref{likehood_function}, $D\left( {{p_{y,0}}\left\| {{p_{y,1}}} \right.} \right)$ and $D\left( {{p_{y,1}}\left\| {{p_{y,0}}} \right.} \right)$ are, respectively, given as
\begin{subequations}
\begin{align}
&D\left( {{p_{y,0}}\left\| {{p_{y,1}}} \right.} \right) =  - \frac{1}{{\ln 2}} - {\mathbb{E}_{{z_{\rm{w}}}}}\left\{ {{{\log }_2}\sum\limits_{k = 1}^K {{p_k}} } \right.\nonumber\\
&\quad \times \left. {\exp \left( { - \frac{{{{\left| {{z_{\rm{w}}} - g_{\rm{w}}^{\rm{*}}{x_k}} \right|}^2}}}{{\sigma _{\rm{w}}^2}}} \right)} \right\}\label{D_p01},\\
&D\left( {{p_{y,1}}\left\| {{p_{y,0}}} \right.} \right) = \sum\limits_{k = 1}^K {{p_k}} {\mathbb{E}_{{z_{\rm{w}}}}}\left\{ {{{\log }_2}\sum\limits_{j = 1}^K {{p_j}} } \right.\nonumber\\
&\quad \left. { \times \exp \left( { - \frac{{{{\left| {g_{\rm{w}}^{\rm{*}}\left( {{x_k} - {x_j}} \right) + {z_{\rm{w}}}} \right|}^2}}}{{\sigma _{\rm{w}}^2}}} \right)} \right\} + \frac{{\left| {g_{\rm{w}}^2{P_{\rm{A}}}} \right|}}{{\left( {\ln 2} \right)\sigma _{\rm{w}}^2}} + \frac{1}{{\ln 2}}.\label{D_p10}
\end{align}
\end{subequations}

Covert communication is achieved  for a given  $\varepsilon$ if the detection error probability
$ \xi $ is no less than $1 - \varepsilon $, i.e.,
\begin{align}
\xi  \ge 1 - \varepsilon, \varepsilon \in \left[ {0,1} \right],
\end{align}
where $\varepsilon $  is a small number determining the required covertness level.

Therefore, to achieve  covert communication with the given $\varepsilon$, i.e., $ \xi    \ge 1 - \varepsilon $, the KL divergences  of the likelihood functions should satisfy  one of the following constraints:
\begin{subequations}\label{Dp0p1}
\begin{align}
&D\left( {{p_{y,0}}\left\| {{p_{y,1}}} \right.} \right)\le 2{\varepsilon ^2},\\
&D\left( {{p_{y,1}}\left\| {{p_{y,0}}} \right.} \right) \le 2{\varepsilon ^2}.
\end{align}
\end{subequations}

{Both} of the above constraints can meet the requirements of covert communication, but these two constraints are not exactly the same, as we discuss next.

\section{Optimal Signaling Design under Covert Constraints}

In this section, we investigate the design of optimal probability of discrete constellation points
 for covert transmission with covertness  constraints $D\left( {{p_{y,0}}\left\| {{p_{y,1}}} \right.} \right) \le 2{\varepsilon ^2}$ or
$D\left( {{p_{y,1}}\left\| {{p_{y,0}}} \right.} \right) \le 2{\varepsilon ^2}$  \cite{Bash13,Wornell16,Bloch16,Yan2019Gaussian}.

\subsection{Case of ${D}\left( {{p_{y,0}}\left\| {{p_{y,1}}} \right.} \right) \le 2{\varepsilon ^2}$}

For the probabilistic constellation shaping scheme,   we aim to  maximize
the achievable rate of Bob ${R_{\rm{b}}}$ by optimizing the distribution  of discrete constellation inputs, while satisfying  both the covert
transmission constraint and the discrete distribution constraint. Mathematically,  the  covert discrete constellation input  optimization problem can be formulated as follows
\begin{subequations}\label{orig_pro 01 1}
\begin{align}
  \mathop { \max }\limits_{\left\{ {{p_k}} \right\}} ~& {R_{\rm{b}}}\\
{\rm{s.t.}}~&D\left( {{p_{y,0}}\left\| {{p_{y,1}}} \right.} \right)\le 2{\varepsilon  ^2},\label{orig1a}\\
&\Pr \left( {X = {x_k}} \right) = {p_k} \ge {\rm{0}},\label{orig1b}\\
& \sum\limits_{k = 1}^K {{p_k}{{\left| {{x_k}} \right|}^2}}  \le {P_\text{A}},\label{orig1c}\\
& \sum\limits_{k = 1}^K {{p_k}}  = 1,k = 1,...,K.\label{orig1d}
\end{align}
\end{subequations}

Since  the KL-divergence $D\left( {{p_{y,0}}\left\| {{p_{y,1}}} \right.} \right)$  in \eqref{D_p10}   is not an analytical expression,   constraint \eqref{orig1a}   is  intractable.
    To circumvent this, we
 derive an explicit  upper bound for the KL-divergence $D\left( {{p_{y,0}}\left\| {{p_{y,1}}} \right.} \right)$, which is given by
 \begin{align}\label{KL_p0p1_ub}
{D_{{\rm{U}}}}\left( {{p_{y,0}}\left\| {{p_{y,1}}} \right.} \right)
= - {\log _2}\sum\limits_{k = 1}^K {{p_k}} \exp \left( { - \frac{{{{\left| {g_{\rm{w}}^{\rm{*}}{x_k}} \right|}^{\rm{2}}}}}{{\sigma _{\rm{w}}^2}}} \right).
\end{align}

{The} details of derivations for \eqref{KL_p0p1_ub}  can be found  in Appendix A.
Based on  \eqref{KL_p0p1_ub}, problem \eqref{orig_pro 01 1} can be rewritten  as
\begin{subequations}\label{orig_2}
\begin{align}
  \mathop { \max }\limits_{\left\{ {{p_k}} \right\}} ~& {R_{\rm{b}}}\\
{\rm{s.t.}}~&- {\log _2}\sum\limits_{k = 1}^K {{p_k}} \exp \left( { - \frac{{{{\left| {g_{\rm{w}}^{\rm{*}}{x_k}} \right|}^{\rm{2}}}}}{{\sigma _{\rm{w}}^2}}} \right) \le 2{\varepsilon  ^2},\\
&\eqref{orig1b}, \eqref{orig1c}, \eqref{orig1d}.\nonumber
\end{align}
\end{subequations}

 In order to put problem \eqref{orig_2} in a more concise form,  we define the following variables \begin{subequations}
\begin{align}
&{\bf{x}} \buildrel \Delta \over = {[{x_1},...,{x_K}]^T},\\
&{\bf{p}} \buildrel \Delta \over = {[{p_1},...,{p_K}]^T},\\
 &{{ {\bf{q}}}} \buildrel \Delta \over ={\left[ {{{\log }_2}{{\bf{p}}^T}{{{\bf{\hat q}}}_1},...,{{\log }_2}{{\bf{p}}^T}{{{\bf{\hat q}}}_K}} \right]^T},\\
&{{\hat {\bf{q}}}_l} \buildrel \Delta \over = \left[ \begin{array}{l}
\exp \left( { - \frac{{{{\left| {g_{\rm{b}}^{\rm{*}}\left( {{x_l} - {x_1}} \right) + {z_{\rm{b}}}} \right|}^2}}}{{\sigma _{\rm{b}}^2}}} \right)\\
~~~~~~~~~~~~~~~...\\
\exp \left( { - \frac{{{{\left| {g_{\rm{b}}^{\rm{*}}\left( {{x_l} - {x_K}} \right) + {z_{\rm{b}}}} \right|}^2}}}{{\sigma _{\rm{b}}^2}}} \right)
\end{array} \right],\forall l \in K,\\
 &{\bf{t}} \buildrel \Delta \over = {\left[ {\exp \left( { - \frac{{{{{\left| {g_{\rm{w}}^{\rm{*}}{x_1}} \right|}}^2}}}{{\sigma _{\rm{w}}^2}}} \right),...,\exp \left( { - \frac{{{{{\left| {g_{\rm{w}}^{\rm{*}}{x_K}} \right|}}^2}}}{{\sigma _{\rm{w}}^2}}} \right)} \right]^T},\\
 &\phi \left( {\bf{p}} \right)\buildrel \Delta \over ={\mathbb{E}_{{z_{\rm{b}}}}}\left\{ {{{\bf{p}}^T}{\bf{q}}} \right\}.
\end{align}
\end{subequations}

{Furthermore},  the rate of Bob  ${R_{\rm{b}}}$ and the upper bound of the KL-divergence ${D_{\rm{U}}}\left( {{p_{0}}||{p_{1}}} \right)$
 can be,  respectively, rewritten as follows
 \begin{subequations}
\begin{align}
 &{R_{\rm{b}}}= - \phi \left( {\bf{p}} \right)  - \frac{1}{{\ln 2}},\\
  &{D_{\rm{U}}}\left( {{p_{y,0}}\left\| {{p_{y,1}}} \right.} \right) =  - {\log _2}{{\bf{p}}^T}{\bf{t}}.
\end{align}
\end{subequations}

Since $\frac{1}{{\ln 2}}$ is constant and maximizing $ - \phi \left( {\bf{p}} \right)$ is equivalent to minimizing $ \phi \left( {\bf{p}} \right)$, problem \eqref{orig_2}  can be reformulated  as
\begin{subequations} \label{pro_01 2}
\begin{align}
\mathop { \min }\limits_{{\bf{p}}} &~ \phi \left( {\bf{p}} \right) \label{obj_01}\\
{\rm{s}}.{\rm{t}}.&-{\log _2}{{\bf{p}}^T}{\bf{t}} \le 2{\varepsilon ^2},\label{pro_01 2a}\\
 &{{\bf{p}}^T}{{\bf{1}}_K} = 1,\label{pro_01 2b}\\
 &{\bf{p}}^T\left( {{\bf{x}}  \odot  {\bf{x}}} \right) \le {P_{\rm{A}}},\label{pro_01 2c}\\
 &{\bf{p}} \ge {\bf{0}},\label{pro_01 2d}
\end{align}
\end{subequations}
where ${{{\bf{1}}_K}}$  is a $K \times 1$ vector with  all elements equal to $1$.

Problem \eqref{pro_01 2} is now a convex  problem, and  we adopt the gradient projection method to solve it. Specifically, let $\nabla \phi \left( {\bf{p}} \right)$ denote the gradient  of the objective function \eqref{obj_01}, which  is given by
\begin{subequations}
\begin{align}
\nabla \phi \left( {\bf{p}} \right) &= {\mathbb{E}_{{z_{\rm{b}}}}}\left\{ {{\bf{q}} + {\bf{Qp}}} \right\}=\int_{ - \infty }^\infty  {{f_{{z_{\rm{b}}}}}\left( {{z_{\rm{b}}}} \right)\left( {{\bf{q}} + {\bf{Qp}}} \right)d} {z_{\rm{b}}}.
\end{align}
\end{subequations}

{Here}, ${f_{{z_{\rm{b}}}}}\left( {{z_{\rm{b}}}} \right){\rm{ = }}\frac{{\rm{1}}}{{\pi \sigma _{\rm{b}}^2}}\exp \left( { - \frac{{{{\left| {{z_{\rm{b}}}} \right|}^2}}}{{\sigma _{\rm{b}}^2}}} \right)$ denotes the probability density function of ${z_{\rm{b}}}$,  ${\bf{Q}}\buildrel \Delta \over =\left[ {{Q_{i,j}}} \right]$, where ${Q_{i,j}} \buildrel \Delta \over = \frac{{{\bf{\hat q}}_j^T{{\bf{e}}_i}}}{{{\bf{\hat q}}_j^T{\bf{p}}\ln 2}}$, and  ${{\bf{e}}_i}$ is the unit vector where the $i$th element is $1$ and the other elements are $0$.

However, neither the objective function $\phi \left( {\bf{p}} \right)$  or the gradient  $\nabla \phi \left( {\bf{p}} \right)$  has  an analytic expression. To tackle this challenge,  we adopt the numerical integration method to approximate $\phi \left( {\bf{p}} \right)$ and $\nabla \phi \left( {\bf{p}} \right)$, i.e.,
\begin{subequations}
\begin{align}
\quad \tilde \phi \left( {\bf{p}} \right) &= \int_{ - {\tau _1} }^{ {\tau _1}} {{f_{{z_{\rm{b}}}}}\left( {{z_{\rm{b}}}} \right)\left( {{{\bf{p}}^T}{\bf{q}}} \right)d} {z_{\rm{b}}}  ,\label{obj_func}\\
\quad \nabla \tilde \phi \left( {\bf{p}} \right) &= \int_{ - {\tau _2} }^{ {\tau _2}}  {{f_{{z_{\rm{b}}}}}\left( {{z_{\rm{b}}}} \right)\left( {{\bf{q}} + {\bf{Qp}}} \right)d} {z_{\rm{b}}}, \label{obj_func_gradient}
\end{align}
\end{subequations}
where $\left[ { - {\tau _1} , {\tau _1}} \right]$ and $\left[ { - {\tau _2} , {\tau _2}} \right]$ denote the integration intervals, by defining  ${\tau _1} > 0$, ${\tau _2} > 0$. Furthermore, $\tilde \phi \left( {\bf{p}} \right)$ and $\nabla \tilde \phi \left( {\bf{p}} \right)$ denote the approximations of the objective function and  its gradient, respectively.

Furthermore, let ${{\bf{p}}_0}$ denote a feasible starting point, and ${{\bf{p}}_n}$ denote the $n$th iteration feasible point. With approximate gradient $ \nabla \tilde \phi \left( {\bf{p}} \right)$,
the   gradient descent iteration step is given as
 \begin{align}{{{\bf{\hat p}}}_{n + 1}} = {{\bf{p}}_n} - {\alpha _n}\nabla \tilde \phi \left( {{\bf{p}}_n} \right),\end{align}
 where ${\alpha _n} \in \left( {0,1} \right]$ is  the stepsize of the $n$th iteration. To choose a proper step size ${\alpha _n}$  with a sufficient decrease, we adopt  the backtracking line search algorithm,   given in Algorithm $1$.

Then, we   project  ${{{{\bf{\hat p}}}_{n + 1}}}$ in the feasible region of problem \eqref{pro_01 2}.
Specifically, when  ${{{{\bf{\hat p}}}_{n + 1}}}$ satisfies constraints \eqref{pro_01 2a}-\eqref{pro_01 2d},  ${{{{\bf{ p}}}_{n + 1}}}={{{{\bf{\hat p}}}_{n + 1}}}$ is in the     feasible region.  Otherwise, we need to find the closest  point ${{{{\bf{ p}}}_{n + 1}}}$ in the feasible region as the  projection of   ${{{{\bf{\hat p}}}_{n + 1}}}$.

Mathematically,    the projection operation of ${{{{\bf{\hat p}}}_{n + 1}}}$  can be formulated as  follows
\begin{subequations}\label{pro_01 3}
\begin{align}
  \mathop {\min }\limits_{{{\bf{p}}_{n + 1}}} &~ {\left\| {{{\bf{p}}_{n + 1}} - {{{\bf{\hat p}}}_{n + 1}}} \right\|^2}\\
{\rm{s}}.{\rm{t}}.&~-{\log _2}{{{\bf{p}}_{n + 1}^T}}{\bf{t}} \le 2{\varepsilon ^2},\\
 &~{{{\bf{p}}_{n + 1}^T}}{{\bf{1}}_K} = 1,\\
 &~{{\bf{p}}_{n + 1}^T}\left( {{\bf{x}}  \odot  {\bf{x}}} \right) \le {P_{\rm{A}}},\\
 &~{{\bf{p}}_{n + 1}} \ge {\bf{0}}.
\end{align}
\end{subequations}

\begin{algorithm}[htb]
	\caption{: Backtracking Line Search for Stepsize ${\alpha _n}$.}
	\begin{algorithmic}[1]
		\State {\bf{Input:}} ${{{\bf{ p}}}_n}$, $\tilde \phi {\left( {{{\bf{p}}_{n }}} \right)}$, $\nabla \tilde \phi \left( {{\bf{p}}_n} \right)$ and $\bar \alpha  > 0$, $\rho ,c \in \left( {0,1} \right)$;
		\State Update ${{{\bf{\hat p}}}_{n+1}}={{\bf{p}}_n} - {\alpha _n}\nabla \tilde \phi \left( {{\bf{p}}_n} \right)$;
		\State {\bf{if}} ${{{\bf{\hat p}}}_{n+1}}$ satisfies constraints \eqref{pro_01 2a}-\eqref{pro_01 2d}
		\State ~~~${{{{\bf{ p}}}_{n + 1}}}={{{{\bf{\hat p}}}_{n + 1}}}$;
		\State  {\bf{else}}
		\State ~~~Solve problem \eqref{pro_01 3} over ${{{\bf{\hat p}}}_{n+1}}$ to obtain ${{{{\bf{ p}}}_{n + 1}}}$;
		\State {\bf{end}};
		\State {\bf{While}}
		$\tilde \phi \left( {{{{\bf{ p}}}_{n + 1}}} \right) \le \tilde \phi \left( {{{\bf{p}}_{n }}} \right) + c\bar \alpha \nabla \tilde \phi {\left( {{{\bf{p}}_{n }}} \right)^T}\left( {{{{{\bf{ p}}}_{n + 1}}} - {{\bf{p}}_{n }}} \right)$;
		\State ~~~$\bar \alpha  \leftarrow  \rho \bar \alpha$;
		\State ~~~Update ${{{\bf{\hat p}}}_{n+1}}={{\bf{p}}_n} - {\alpha _n}\nabla \tilde \phi \left( {{\bf{p}}_n} \right)$;
		\State ~~~{\bf{if}} ${{{\bf{\hat p}}}_{n+1}}$ satisfies constraints \eqref{pro_01 2a}-\eqref{pro_01 2d}
		\State ~~~~~~${{{{\bf{ p}}}_{n + 1}}}={{{{\bf{\hat p}}}_{n + 1}}}$;
		\State  ~~~{\bf{else}}
		\State ~~~~~~solve problem \eqref{pro_01 3} over ${{{\bf{\hat p}}}_{n+1}}$ to obtain ${{{{\bf{ p}}}_{n + 1}}}$;
		\State ~~~{\bf{end}};
		\State   {\bf{end}};
		\State   {\bf{return}}  ${\alpha _{n }} = \bar \alpha $.
	\end{algorithmic}
\end{algorithm}

%
%
Therefore, we propose   the approximate gradient descent  projection method to
 efficiently solve problem \eqref{pro_01 2}, which is summarized
in Algorithm 2.
 \begin{algorithm}[htb]
 	\caption{ Inexact Gradient Descent Projection Method.}
 	\begin{algorithmic}[1]
 		\State {\bf{Input}}:
 		\parbox[t]{\dimexpr\linewidth-\algorithmicindent - 0.45cm}{choose $K \ge 2$ and choose a random starting point ${{\bf{p}}_0}$ which satisfies constraints \eqref{pro_01 2a}-\eqref{pro_01 2d}, set ${c_2}$ as the stopping parameter and $n=0$;\strut}			
 		\State {\bf{Repeat}}
 		\State ~~~Let $n \leftarrow n + 1$;
 		\State ~~~Update $\tilde \phi \left( {{{\bf{p}}_{n - 1}}} \right)= \int_{ - {\tau _1} }^{ {\tau _1}} {{f_{{z_{\rm{b}}}}}\left( {{z_{\rm{b}}}} \right)\left( {{{{\bf{p}}_{n - 1}^T}}{\bf{q}}} \right)d} {z_{\rm{b}}}$;
 		\State ~~~Update ${\nabla \tilde \phi \left( {{{\bf{p}}_{n - 1}}} \right)}= \int_{ - {\tau _2} }^{ {\tau _2}}  {{f_{{z_{\rm{b}}}}}\left( {{z_{\rm{b}}}} \right)\left( {{\bf{q}} + {\bf{Q}}{{\bf{p}}_{n - 1}}} \right)d} {z_{\rm{b}}}$;
 		\State  ~~~Compute stepsize  ${\alpha _{n - 1}}$ by Algorithm $1$;
 		\State ~~~Update ${{{\bf{\hat p}}}_{n}}={{\bf{p}}_{n-1}} - {\alpha _{n-1}}\nabla \tilde \phi \left( {{\bf{p}}_{n-1}} \right)$;
 		\State ~~~{\bf{if}} ${{{\bf{\hat p}}}_{n}}$ satisfies constraints \eqref{pro_01 2a}-\eqref{pro_01 2d}
 		\State ~~~~~~${{{{\bf{ p}}}_{n }}}={{{{\bf{\hat p}}}_{n }}}$;
 		\State ~~~{\bf{else}}
 		\State ~~~~~~solve problem \eqref{pro_01 3} over ${{{\bf{\hat p}}}_{n}}$ to obtain ${{{{\bf{ p}}}_{n }}}$;
 		\State ~~~{\bf{end}};
 		\State {\bf{Until}} $\left\| {{{\bf{p}}_n} - {{\bf{p}}_{n - 1}}} \right\| \le {c_2}$;
 		\State   {\bf{Output}} ${{\bf{P}}^{{\rm{opt}}}} = {{\bf{p}}_n}$.
 	\end{algorithmic}
 \end{algorithm}

\subsection{Case of ${D}\left( {{p_{y,1}}\left\| {{p_{y,0}}} \right.} \right) \le 2{\varepsilon ^2}$}

In this subsection, we further consider the other covert constraint ${D}\left( {{p_{y,1}}\left\| {{p_{y,0}}} \right.} \right) \le 2{\varepsilon ^2}$, and  the corresponding covert rate  optimization problem can be formulated as
\begin{subequations}\label{orig_pro 10 1}
\begin{align}
  \mathop { \max }\limits_{\left\{ {{p_k}} \right\}} ~& {R_{\rm{b}}}\\
{\rm{s.t.}}~&{D}\left( {{p_{y,1}}\left\| {{p_{y,0}}} \right.} \right) \le 2{\varepsilon  ^2},\label{orig2a}\\
&\Pr \left( {X = {x_k}} \right) = {p_k} \ge {\rm{0}},\label{orig2b}\\
& \sum\limits_{k = 1}^K {{p_k}{{\left| {{x_k}} \right|}^2}}  \le {P_{\rm{A}}},\label{orig2c}\\
& \sum\limits_{k = 1}^K {{p_k}}  = 1,k = 1,...,K.\label{orig2d}
\end{align}
\end{subequations}

To handle intractable constraint \eqref{orig2a}, we first  derive    an upper bound    on $D\left( {{p_{y,1}}\left\| {{p_{y,0}}} \right.} \right)$ denoted by ${D_{{\rm{U}}}}\left( {{p_{y,1}}\left\| {{p_{y,0}}} \right.} \right)$, which  is given by
\begin{align}\label{KL_p1p0_ub}
{D_{{\rm{U}}}}\left( {{p_{y,1}}\left\| {{p_{y,0}}} \right.} \right)
=&\sum\limits_{k = 1}^K {{p_k}} {\log _2}\sum\limits_{j = 1}^K {{p_j}}\exp \left( { - \frac{{{{\left| {g_{\rm{w}}^{\rm{*}}}\left( {{x_k} - {x_j}} \right) \right|}^{\rm{2}}}}}{{2\sigma _{\rm{w}}^2}}} \right)\nonumber\\
&+ \frac{1}{{\ln 2}} + \frac{\left|{{ {{g_{\rm{w}}^2}}}{P_{\rm{A}}}}\right|}{{\left( {\ln 2} \right)\sigma _{\rm{w}}^2}}-1.
\end{align}

{The} details of derivations for \eqref{KL_p1p0_ub} are given in Appendix B.
Then,   problem \eqref{orig_pro 10 1} can be reformulated as
\begin{subequations}\label{orig_3}
\begin{align}
  \mathop { \max }\limits_{\left\{ {{p_k}} \right\}} ~& {R_{\rm{b}}}\\
{\rm{s.t.}}~&{D_{{\rm{U}}}}\left( {{p_{y,1}}\left\| {{p_{y,0}}} \right.} \right)\le 2{\varepsilon  ^2},\label{orig_3a}\\
&\eqref{orig2b}, \eqref{orig2c}, \eqref{orig2d}.\nonumber
\end{align}
\end{subequations}

By defining the following equations to simplify   ${D_{{\rm{U}}}}\left( {{p_{y,1}}\left\| {{p_{y,0}}} \right.} \right)$
\begin{subequations}
\begin{align}
 &{{\bf{s}}_{{\rm{w}},k}} \buildrel \Delta \over = \left[ \begin{array}{l}
\exp \left( { - \frac{{{{\left| {g_{\rm{w}}^{\rm{*}}\left( {{x_k} - {x_1}} \right)} \right|}^2}}}{{2\sigma _{\rm{w}}^2}}} \right),...,
\exp \left( { - \frac{{{{\left| {g_{\rm{w}}^{\rm{*}}\left( {{x_k} - {x_K}} \right)} \right|}^2}}}{{2\sigma _{\rm{w}}^2}}} \right)
\end{array} \right]^T,\\
 &{{\bf{v}}_{\rm{w}}}\left( {\bf{p}} \right)\buildrel \Delta \over = {\left[ {{{\log }_2}{{\bf{p}}^T}{{\bf{s}}_{{\rm{w}},1}},...,{{\log }_2}{{\bf{p}}^T}{{\bf{s}}_{{\rm{w}},K}}} \right]^T},
\end{align}
\end{subequations}
we     obtain
\begin{align}
  &{D_{\rm{U}}}\left( {{p_{y,1}}\left\| {{p_{y,0}}} \right.} \right) = {{\bf{p}}^T}{{\bf{v}}_{\rm{w}}}\left( {\bf{p}} \right) + \frac{1}{{\ln 2}} + \frac{\left|{{ {{g_{\rm{w}}^2}}}{P_x}}\right|}{{\left( {\ln 2} \right)\sigma _{\rm{w}}^2}}-1.
\end{align}

Unfortunately,  ${D_{\rm{U}}}\left( {{p_{y,1}}\left\| {{p_{y,0}}} \right.} \right)$ is non-convex in ${\bf{p}}$, and the covert constraint is also non-convex.
To handle this issue, we   apply the first order Taylor expansion to ${D_{\rm{U}}}\left( {{p_{y,1}}\left\| {{p_{y,0}}} \right.} \right)$. Specifically,
the derivative of ${D_{\rm{U}}}\left( {{p_{y,1}}\left\| {{p_{y,0}}} \right.} \right)$ at  ${{{{\bf{\bar p}}}_n}}$ is given by
\begin{align}
\nabla {D_{\rm{U}}}\left( {{p_{y,1}}\left\| {{p_{y,0}}} \right.} \right)\left| {_{{\bf{p}} = {{{{\bf{\bar p}}}_n}}}} \right.
= {{\bf{v}}_{\rm{w}}}\left( {{{\bf{\bar p}}}_n} \right) + \nabla {{\bf{v}}_{\rm{w}}}{{{\bf{\bar p}}}_n},
\end{align}
where  $\nabla {\bf{v}}_{\rm{w}} = {\left[ {\frac{{{{\bf{s}}_{{\rm{w}},1}}}}{{{{{\bf{\bar p}}}_n^T}{{\bf{s}}_{{\rm{w}},1}}}},...,\frac{{{{\bf{s}}_{{\rm{w}},K}}}}{{{{{\bf{\bar p}}}_n^T}{{\bf{s}}_{{\rm{w}},K}}}}} \right]_{K \times K}}$. Then, the first order Taylor expansion of ${{\bf{p}}^T}{{\bf{v}}_{\rm{w}}}\left( {\bf{p}} \right)$ is given as follows
\begin{align}
L\left( {\bf{p}} \right)\approx {\bf{\bar p}}_n^T{\bf{v}}_{\rm{w}}\left( {{{\bf{\bar p}}}_n} \right) + {\left( {{\bf{v}}_{\rm{w}}\left( {{{\bf{\bar p}}}_n} \right) + \nabla {\bf{v}}_{\rm{w}}{{{\bf{\bar p}}}_n}} \right)^T}\left( {{\bf{p}} - {{{\bf{\bar p}}}_n}} \right).
\end{align}

Then,    constraint \eqref{orig_3a} can be  recast to a convex form as
\begin{align}
L\left( {\bf{p}} \right) + \frac{1}{{\ln 2}} + \frac{\left|{{ {{g_{\rm{w}}^2}}}{P_{\rm{A}}}}\right|}{{\left( {\ln 2} \right)\sigma _{\rm{w}}^2}}-1 \le 2{\varepsilon ^2}\label{constraint_1_2}.
\end{align}

Thus,   problem \eqref{orig_pro 10 1} can be reformulated as
\begin{subequations}\label{pro_10 2}
\begin{align}
  \mathop {\min }\limits_{{{\bf{p}}}} &~ \phi \left( {\bf{p}} \right) \\
\quad {\rm{s}}{\rm{.t}}{\rm{.}} &~L\left( {\bf{p}} \right) + \frac{1}{{\ln 2}} + \frac{\left|{{ {{g_{\rm{w}}^2}}}{P_{\rm{A}}}}\right|}{{\left( {\ln 2} \right)\sigma _{\rm{w}}^2}}-1 \le 2{\varepsilon ^2},\\
\quad &~ {\bf{p}}^T{{\bf{1}}_K} = 1,\\
\quad & ~{\bf{p}}^T\left( {{\bf{x}} \odot {\bf{x}}} \right) \le {P_{\rm{A}}},\\
\quad & ~{{\bf{p}}} \ge 0,
\end{align}
\end{subequations}
which is convex.

 Similarly, problem \eqref{pro_10 2} can be efficiently solved by the inexact gradient descent projection method. The details are omitted due to  space limitation.

\section{Signaling Design with Approximate Covert Rate Expression}


Due to   the expectation operation,
the achievable rate in \eqref{MI}  does not have a closed-form expression, and can only be computed numerically using the approximate gradient descent method at the expense of high computational complexity.
  To strike a balance
between complexity and performance, we further derive analytical upper bound and lower bound on the  achievable rate in \eqref{MI}.

\textbf{Lemma 1}: An upper bound    ${R_{\rm{b}}^{\text{U}}}$ on the rate ${R_{\rm{b}}}$  is given by
\begin{align}\label{mutual_I ub}
{R_{\rm{b}}^{\text{U}}}
=- \sum\limits_{k = 1}^K {{p_k}} {\log _2}\sum\limits_{j = 1}^K {{p_j}} \exp \left( { - \frac{{{{\left| {{g_{\rm{b}}^{\rm{*}}}\left( {{x_k} - {x_j}} \right)} \right|}^2}}}{{\sigma _{\rm{b}}^2}}} \right),
\end{align}
while a lower bound ${R_{\rm{b}}^{\text{L}}}$ on the rate ${R_{\rm{b}}}$ is given as
\begin{align}\label{mutual_I lb}
{R_{\rm{b}}^{\text{L}}}
=& - \sum\limits_{k = 1}^K {{p_k}} {\log _2}\sum\limits_{j = 1}^K {{p_j}}\exp \left( { - \frac{{{{\left| {g_{\rm{b}}^{\rm{*}}}\left( {{x_k} - {x_j}} \right) \right|}^{\rm{2}}}}}{{2\sigma _{\rm{b}}^2}}} \right)\nonumber\\
&- \frac{1}{{\ln 2}}+1.
\end{align}
Please find the derivation in Appendices C and D.

In this section, we adopt the upper bound and lower bound on the achievable rate ${R_{\rm{b}}}$ in our following analysis.

\subsection{Maximizing  ${R_{\rm{b}}^{\text{U}}}$ }

In this subsection, we  consider the upper bound on the achievable rate for Bob ${R_{\rm{b}}^{\text{U}}}$ as the objective function to find the optimal probability of discrete constellation points set. Specifically, we study  beamforming design with the objective of maximizing ${R_{\rm{b}}^{\text{U}}}$, subject to the covert transmission
constraint, and the discrete constellation set.

\subsubsection{${D_{\rm{U}}}\left( {{p_{y,0}}\left\| {{p_{y,1}}} \right.} \right) \le 2{\varepsilon ^2}$}

Finding the optimal probability of discrete constellation set  can be equivalently  written  as the  following optimization problem
\begin{subequations}\label{uper_pro 01 2}
\begin{align}
 \mathop { \max }\limits_{\left\{ {{p_k}} \right\}} &~  {R_{\rm{b}}^{\text{U}}}\\
{\rm{s}}{\rm{.t}}.&~{D_{\rm{U}}}\left( {{p_{y,0}}\left\| {{p_{y,1}}} \right.} \right) \le 2{\varepsilon  ^2},\\
&~\Pr \left( {X = {x_k}} \right) = {p_k} \ge {\rm{0}},\label{up1b}\\
&~ \sum\limits_{k = 1}^K {{p_k}{{\left| {{x_k}} \right|}^2}}  \le {P_{\rm{A}}},\label{up1c}\\
&~ \sum\limits_{k = 1}^K {{p_k}}  = 1,k = 1,...,K.\label{up1d}
\end{align}
\end{subequations}

In order to solve problem \eqref{uper_pro 01 2}, we first define the following variables
\begin{subequations}
\begin{align}
 &{{\bf{r}}_k} \buildrel \Delta \over =  \left[ \begin{array}{l}
 \exp \left( { - \frac{{{{\left| {{g_{\rm{b}}^{\rm{*}}}\left( {{x_k} - {x_1}} \right)} \right|}^2}}}{{\sigma _{\rm{b}}^2}}} \right),...,
 \exp \left( { - \frac{{{{\left| {{g_{\rm{b}}^{\rm{*}}}\left( {{x_k} - {x_K}} \right)} \right|}^2}}}{{\sigma _{\rm{b}}^2}}} \right)
 \end{array} \right]^T,\\
 &{\bf{u}}\left( {\bf{p}} \right) \buildrel \Delta \over = {\left[ {{{\log }_2}{{\bf{p}}^T}{{\bf{r}}_1},...,{{\log }_2}{{\bf{p}}^T}{{\bf{r}}_K}} \right]^T}.
\end{align}
\end{subequations}

{In} this case, we  can  obtain
\begin{align}
  &{R_{\rm{b}}^{\text{U}}} =  - {{\bf{p}}^T}{\bf{u}}\left( {\bf{p}} \right).
\end{align}

Therefore, problem \eqref{uper_pro 01 2} can be reformulated as follows
\begin{subequations}\label{approxi_pro 01}
\begin{align}
 \mathop { \min }\limits_{{\bf{p}}} &~ {f_{\rm{U}}}\left( {\bf{p}} \right) \label{obj_01 1}\\
{\rm{s}}{\rm{.t}}.&~- {\log _2}{{\bf{p}}^T}{\bf{t}} \le 2{\varepsilon ^2},\label{approxi_pro 01a}\\
 &~{{\bf{p}}^T}{{\bf{1}}_K} = 1,\label{approxi_pro 01b}\\
 &~{\bf{p}}^T\left( {{\bf{x}}  \odot  {\bf{x}}} \right) \le {P_{\rm{A}}},\label{approxi_pro 01c}\\
 &~{\bf{p}} \ge {\bf{0}},\label{approxi_pro 01d}
\end{align}
\end{subequations}
where ${f_{\rm{U}}}\left( {\bf{p}} \right)={{{\bf{p}}^T}{\bf{u}}}\left( {\bf{p}} \right)$.
Since the Frank-Wolf method is an algorithm for solving linearly-constrained problems, it makes a linear approximation of the objective function, obtains the feasible descending direction by solving the linear programming, and conducts a one-dimensional search in the feasible region along this direction.
Therefore, we will apply the Frank-Wolf method to solve the optimization problem.

 We use Taylor's expansion to make a linear approximation of the objective function ${f_{\rm{U}}}\left( {\bf{p}} \right)$.
 The first order Taylor expansion  at ${{{\bf{p}}_n}}$ is as follows
\begin{subequations}
\begin{align}
&{f_{\rm{U}}}\left( {\bf{p}} \right) \approx {\bf{p}}_n^T{\bf{u}}\left( {{\bf{p}}_n} \right) + \nabla {f_{\rm{U}}}{\left( {{{\bf{p}}_i}} \right)^T}\left( {{\bf{p}} - {{\bf{p}}_n}} \right),\\
&\nabla {f_{\rm{U}}}\left( {{{\bf{p}}_n}}\right) = {\bf{u}}\left( {{\bf{p}}_n} \right) + \nabla {\bf{u}}{{\bf{p}}_n},
\end{align}
\end{subequations}
where $\nabla {\bf{u}} = {\left[ {\frac{{{{\bf{r}}_{1}}}}{{{{\bf{p}}_n^T}{{\bf{r}}_{1}}}},...,\frac{{{{\bf{r}}_{K}}}}{{{{\bf{p}}_n^T}{{\bf{r}}_{K}}}}} \right]_{K \times K}}$,  and ${{\bf{p}}_n}$ denotes the current iteration point.
Then, we reformulate the  optimization problem of \eqref{approxi_pro 01} as follows
\begin{align}\label{pro_U 01}
 \mathop { \min }\limits_{{\bf{p}}}& ~
 \nabla {f_{\rm{U}}}{\left( {{{\bf{p}}_n}} \right)^T}{\bf{p}} \\
{\rm{s}}{\rm{.t}}.&~\eqref{approxi_pro 01a}, \eqref{approxi_pro 01b}, \eqref{approxi_pro 01c}, \eqref{approxi_pro 01d}\notag.
\end{align}

By applying the Frank-Wolf method, the detailed procedures for solving (40) are summarized in Algorithm $3$. Note that ${\lambda _n}$ is the stepsize of the $n$th iteration and ${{\bf{d}}_n}$ denotes the feasible descending
direction of the $n$th iteration.
\begin{algorithm}[htb]
  \caption{: Solving \eqref{pro_U 01} by Frank-Wolf method.}
  \begin{algorithmic}[1]
    \State  {\bf{Initialization:}} Choose a feasible starting point ${{\bf{p}}_0}$, set $\delta > 0 $ as the stopping parameter, let $n=0$;
    \State {\bf{While}} $\left\| {\nabla {f}{{\left( {{{\bf{p}}_n}} \right)}^T}{{\bf{d}}_n}} \right\| \le \delta $;

    \State  ~~~Solve the linear programming problems \eqref{pro_U 01}
               and  obtain  optimal solution ${{\bf{\bar p}}_n}$ ;
    \State ~~~Construct the feasible descending direction ${{\bf{d}}_n} = {\bf{\bar p}}_n - {\bf{p}}_n$;
    \State ~~~Obtain optimal solution ${\lambda _n}=\mathop {\arg \min }\limits_{0 \le \lambda  \le 1} {f}\left( {{{\bf{p}}_n} + \lambda {{\bf{d}}_n}} \right)$;
    \State ~~~Let ${{\bf{p}}_{n + 1}} = {{\bf{p}}_n} + {\lambda _n}{{\bf{d}}_n}$, $n \leftarrow n + 1;$
    \State {\bf{end}}
    \State   {\bf{Output}} ${{\bf{p}}_n}$
  \end{algorithmic}
\end{algorithm}

\subsubsection{${D_{\rm{U}}}\left( {{p_{y,1}}\left\| {{p_{y,0}}} \right.} \right) \le 2{\varepsilon  ^2}$}
Furthermore, we consider the optimal probabilistic constellation shaping  for covert communications with covert constraint ${D_{\rm{U}}}\left( {{p_{y,1}}\left\| {{p_{y,0}}} \right.} \right) \le 2{\varepsilon  ^2}$, such that
\begin{subequations}\label{uper_pro 10 2}
\begin{align}
 \mathop { \max }\limits_{\left\{ {{p_k}} \right\}}&~    {R_{\rm{b}}^{\text{U}}}\\
{\rm{s}}{\rm{.t}}.&~{D_{\rm{U}}}\left( {{p_{y,1}}\left\| {{p_{y,0}}} \right.} \right) \le 2{\varepsilon  ^2},\label{uper_pro 10 2a}\\
&~\eqref{up1b}, \eqref{up1c}, \eqref{up1d}\notag,
\end{align}
\end{subequations}
 which is non-convex.

 Similar to problem \eqref{orig_3},  by replacing   constraint \eqref{uper_pro 10 2a} by constraint \eqref{constraint_1_2}, we can obtain the optimization problem
\begin{subequations} \label{pro_U 10}
\begin{align}
 \mathop { \min }\limits_{{\bf{p}}}&~\nabla {f_U}{\left( {{{\bf{p}}_n}} \right)^T}{\bf{p}} \\
{\rm{s}}{\rm{.t}}.& ~L\left( {\bf{p}} \right) + \frac{1}{{\ln 2}} + \frac{\left|{{ {{g_{\rm{w}}^2}}}{P_{\rm{A}}}}\right|}{{\left( {\ln 2} \right)\sigma _{\rm{w}}^2}}-1 \le 2{\varepsilon ^2},\\
 &~\eqref{approxi_pro 01a}, \eqref{approxi_pro 01b}, \eqref{approxi_pro 01c}, \eqref{approxi_pro 01d}\notag.
\end{align}
\end{subequations}

{Then}, we  apply the Frank-Wolf method  to solve problem \eqref{uper_pro 01 2}, and the details are omitted since the corresponding algorithm is
similar to Algorithm 3.

\subsection{Maximizing $ {R_{\rm{b}}^{\text{L}}}$}

In this subsection, we further study the lower bound beamforming design for covert communication by maximizing the lower bound ${R_{\rm{b}}^{\text{L}}}$, while satisfying the covert transmission requirement and the discrete constellation set with $K$.

\subsubsection{${D_{\rm{U}}}\left( {{p_{y,0}}\left\| {{p_{y,1}}} \right.} \right) \le 2{\varepsilon  ^2}$}
 Under the covert constraint ${D_{\rm{U}}}\left( {{p_{y,0}}\left\| {{p_{y,1}}} \right.} \right)\le 2{\varepsilon  ^2}$,
 the optimal probabilistic constellation shaping for covert communications is formulated as follows
\begin{subequations}\label{low_pro 01 1}
\begin{align}
 \mathop { \max }\limits_{\left\{ {{p_k}} \right\}} &~  {R_{\rm{b}}^{\text{L}}}\\
{\rm{s}}{\rm{.t}}.&~{D_{\rm{U}}}\left( {{p_{y,0}}\left\| {{p_{y,1}}} \right.} \right) \le 2{\varepsilon  ^2},\label{lowa}\\
&~\Pr \left( {X = {x_k}} \right) = {p_k} \ge {\rm{0}},\label{lowb}\\
&~ \sum\limits_{k = 1}^K {{p_k}{{\left| {{x_k}} \right|}^2}}  \le {P_{\rm{A}}},\label{lowc}\\
&~ \sum\limits_{k = 1}^K {{p_k}}  = 1,k = 1,...,K.\label{lowd}
\end{align}
\end{subequations}

To solve the problem, we define the following equations
\begin{subequations}
\begin{align}
 &{{\bf{s}}_{{\rm{b}},k}} \buildrel \Delta \over = \left[ \begin{array}{l}
 \exp \left( { - \frac{{{{\left| {g_{\rm{b}}^{\rm{*}}}\left( {{x_k} - {x_1}} \right) \right|}^{\rm{2}}}}}{{2\sigma _{\rm{b}}^2}}} \right),...,
 \exp \left( { - \frac{{{{\left| {g_{\rm{b}}^{\rm{*}}}\left( {{x_k} - {x_K}} \right) \right|}^{\rm{2}}}}}{{2\sigma _{\rm{b}}^2}}} \right)
  \end{array} \right]^T,\\
 &{{\bf{v}}_{\rm{b}}}\left( {\bf{p}} \right) \buildrel \Delta \over = {\left[ {{{\log }_2}{{\bf{p}}^T}{{\bf{s}}_{{\rm{b}},1}},...,{{\log }_2}{{\bf{p}}^T}{{\bf{s}}_{{\rm{b}},K}}} \right]^T},
\end{align}
\end{subequations}
and then  the lower bound   ${R_{\rm{b}}^{\text{L}}}$ can be transformed to
\begin{align}
  &{R_{\rm{b}}^{\text{L}}} =  - {{\bf{p}}^T}{{\bf{v}}_{\rm{b}}}\left( {\bf{p}} \right)- \frac{1}{{\ln 2}}+1.
\end{align}

Thus,  the covert optimization problem is recast  as follows
\begin{subequations}\label{low_pro 01 2}
\begin{align}
 \mathop { \max }\limits_{\bf{p}} &  - {{\bf{p}}^T}{{\bf{v}}_{\rm{b}}}\left( {\bf{p}} \right)\\
{\rm{s}}{\rm{.t}}. &~- {\log _2}{{\bf{p}}^T}{\bf{t}} \le 2{\varepsilon ^2},\label{low02a}\\
 &~{{\bf{p}}^T}{{\bf{1}}_K} = 1,\label{low02b}\\
 &~{\bf{p}}^T\left( {{\bf{x}}  \odot  {\bf{x}}} \right) \le {P_{\rm{A}}},\label{low02c}\\
 &~{\bf{p}} \ge {\bf{0}},\label{low02d}
\end{align}
\end{subequations}

{Similar} to problem \eqref{uper_pro 01 2},  we can apply the Frank-Wolf method to solve problem \eqref{low_pro 01 1}, and the details are omitted.

\subsubsection{${D_{\rm{U}}}\left( {{p_{y,1}}\left\| {{p_{y,0}}} \right.} \right)\le 2{\varepsilon  ^2}$}
 With the covert constraint ${D_{\rm{U}}}\left( {{p_{y,1}}\left\| {{p_{y,0}}} \right.} \right)\le 2{\varepsilon  ^2}$,
 the optimal probabilistic constellation shaping for covert communications is given as
\begin{subequations}\label{low_pro 10 1}
\begin{align}
 \mathop { \max }\limits_{\left\{ {{p_k}} \right\}}  &~ {R_{\rm{b}}^{\text{L}}}\\
{\rm{s}}{\rm{.t}}.&{D_{\rm{U}}}\left( {{p_{y,1}}\left\| {{p_{y,0}}} \right.} \right) \le 2{\varepsilon  ^2},\label{low10a}\\
&\eqref{lowb}, \eqref{lowc}, \eqref{lowd}\notag.
\end{align}
\end{subequations}

By replacing the constraint \eqref{low10a} with \eqref{constraint_1_2}, we   obtain the optimization problem
\begin{subequations}\label{low_pro 10 2}
\begin{align}
\mathop { \max }\limits_{\bf{p}} & ~ - {{\bf{p}}^T}{{\bf{v}}_{\rm{b}}}\left( {\bf{p}} \right)\\
{\rm{s}}{\rm{.t}}.&~L\left( {\bf{p}} \right) + \frac{1}{{\ln 2}} + \frac{\left|{{ {{g_{\rm{w}}^2}}}{P_{\rm{A}}}}\right|}{{\left( {\ln 2} \right)\sigma _{\rm{w}}^2}}-1 \le 2{\varepsilon ^2},\\
&~\eqref{low02b}, \eqref{low02c}, \eqref{low02d}\notag.
\end{align}
\end{subequations}

{Similar} to problem \eqref{uper_pro 10 2},  we can apply a similar method to solve problem \eqref{low_pro 10 1}.

\section{Numerical Results}
In this section, we present and discuss numerical results to assess the performance of the proposed probabilistic constellation shaping  designs. In our simulations,  the discrete constellation input is QAM modulation,
  the total transmit power of Alice
is   ${P_{\rm{A}}} = 10\rm{W}$,
the noise variance of  Willie is  $\sigma _{\rm{w}}^2=1\rm{W}$, and the noise variance of  Bob is $\sigma _{\rm{b}}^{\rm{2}}{\rm{ = }}\frac{{{P_{\rm{A}}}}}{{{{10}^{{{{\rm{SNR}}} \mathord{\left/
 {\vphantom {{{\rm{SNR}}} {10}}} \right.
 \kern-\nulldelimiterspace} {10}}}}}}$. Moreover, we assume that all channels experience Rayleigh flat fading, and $\sigma _1=\sigma _2=1$ \cite{Shahzad2017Covert}.

We first compare  the proposed optimal probabilistic constellation shaping  design with the equiprobable design, starting from the empirical CDF of KL divergence and the rate comparison.

\begin{figure}
    \begin{minipage}[htbp]{0.45\textwidth}
      \centering
      \includegraphics[height=7.5cm,width=7.5cm]{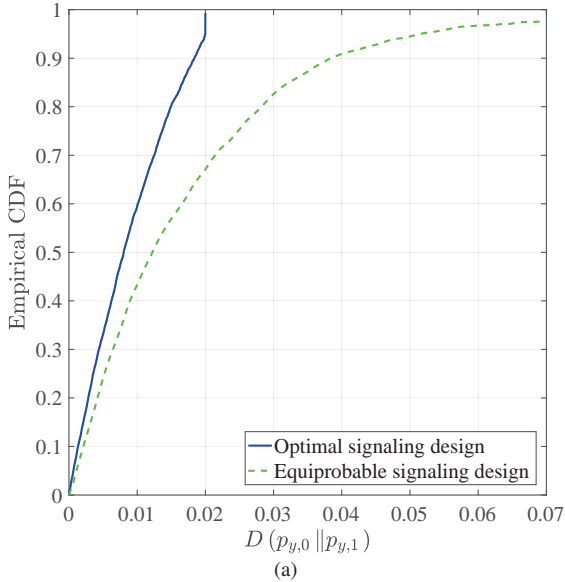}
      \vskip-0.2cm\centering {\footnotesize (a)}
    \end{minipage}
     \begin{minipage}[htbp]{0.45\textwidth}
      \centering
      \includegraphics[height=7.5cm,width=7.5cm]{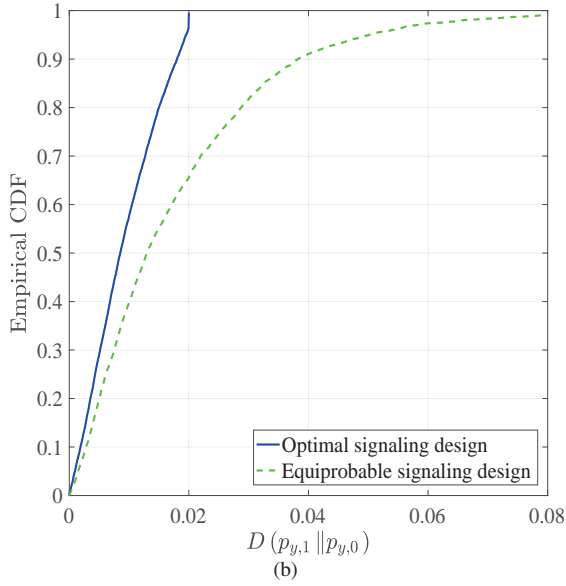}
      \vskip-0.2cm\centering {\footnotesize (b)}
    \end{minipage}\hfill
 \caption{ The empirical CDF of a) $D\left( {{p_{y,0}}\left\| {{p_{y,1}}} \right.} \right)$ and (b) $D\left( {{p_{y,1}}\left\| {{p_{y,0}}} \right.} \right)$ with the covertness threshold   $2{\varepsilon ^2} = 0.02$ for the proposed optimal probabilistic constellation shaping design and the equiprobable design. }
 \label{1}  
\end{figure}
Fig. \ref{1} shows the empirical CDF of the achieved $D\left( {{p_{y,0}}\left\| {{p_{y,1}}} \right.} \right)$ and $D\left( {{p_{y,1}}\left\| {{p_{y,0}}} \right.} \right)$, respectively, for both the proposed optimal  probabilistic constellation shaping design and the equiprobable design with ${\rm{SNR}}=10\rm{dB}$ and $K=8$, where the covertness threshold is
$2{\varepsilon ^2} = 0.02$, i.e., $D\left( {{p_{y,0}}\left\| {{p_{y,1}}} \right.} \right) \le 0.02$ and $D\left( {{p_{y,1}}\left\| {{p_{y,0}}} \right.} \right) \le 0.02$.
As  observed
from Fig. \ref{1}, the proposed optimal  probabilistic constellation shaping design satisfies
the covertness constraint. On the other hand, the equiprobable design
cannot satisfy the covertness constraints.

\begin{figure}
    \begin{minipage}[htbp]{0.45\textwidth}
      \centering
      \includegraphics[height=7.5cm,width=7.5cm]{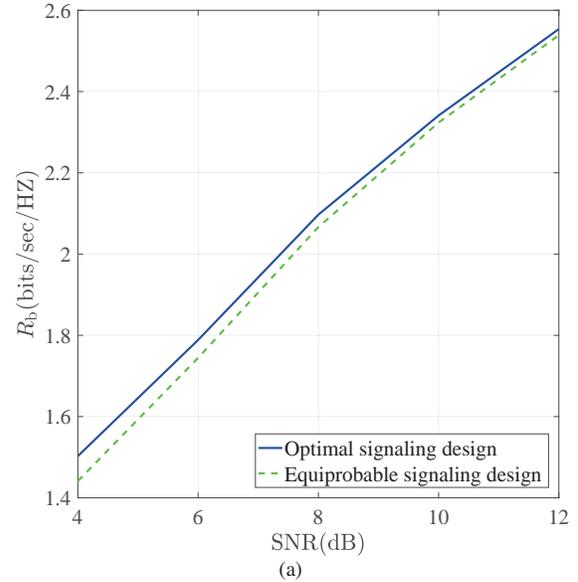}
      \vskip-0.2cm\centering {\footnotesize (a)}
    \end{minipage}
     \begin{minipage}[htbp]{0.45\textwidth}
      \centering
      \includegraphics[height=7.5cm,width=7.5cm]{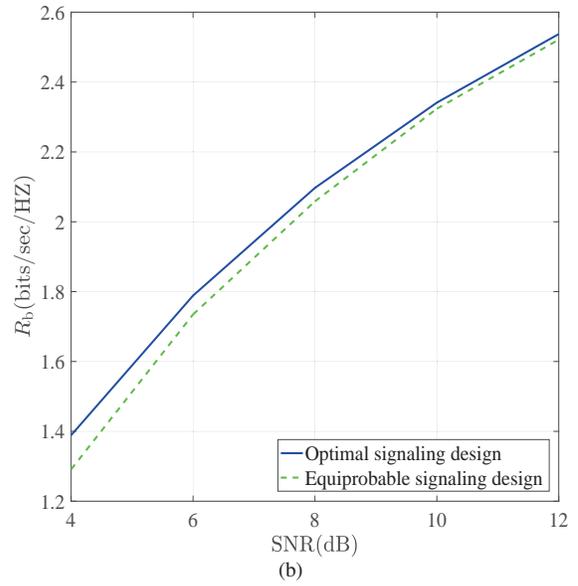}
      \vskip-0.2cm\centering {\footnotesize (b)}
    \end{minipage}\hfill
 \caption{ The  covert rate of a) $D\left( {{p_{y,0}}\left\| {{p_{y,1}}} \right.} \right)$ and (b) $D\left( {{p_{y,1}}\left\| {{p_{y,0}}} \right.} \right)$  for the proposed optimal probabilistic constellation shaping design and the equiprobable design.  }
 \label{2}  
\end{figure}
 Fig. \ref{2} (a) and (b)  show the achievable covert rate of Bob   ${R_{\rm{b}}}$ for the proposed optimal probabilistic constellation shaping design and the equiprobable design with $D\left( {{p_{y,0}}\left\| {{p_{y,1}}} \right.} \right)$ and $D\left( {{p_{y,1}}\left\| {{p_{y,0}}} \right.} \right)$, respectively. We observe that  the optimal probabilistic constellation shaping design is superior when the signal-to-noise ratio is low. Therefore, in practical applications, the proposed optimal  probabilistic constellation shaping design is advantageous in medium and low SNR, and the equiprobable design is suitable for high SNR.


Next, we evaluate the performance of the proposed optimal  probabilistic constellation shaping design.

\begin{figure}
    \begin{minipage}[htbp]{0.45\textwidth}
      \centering
      \includegraphics[height=7.5cm,width=7.5cm]{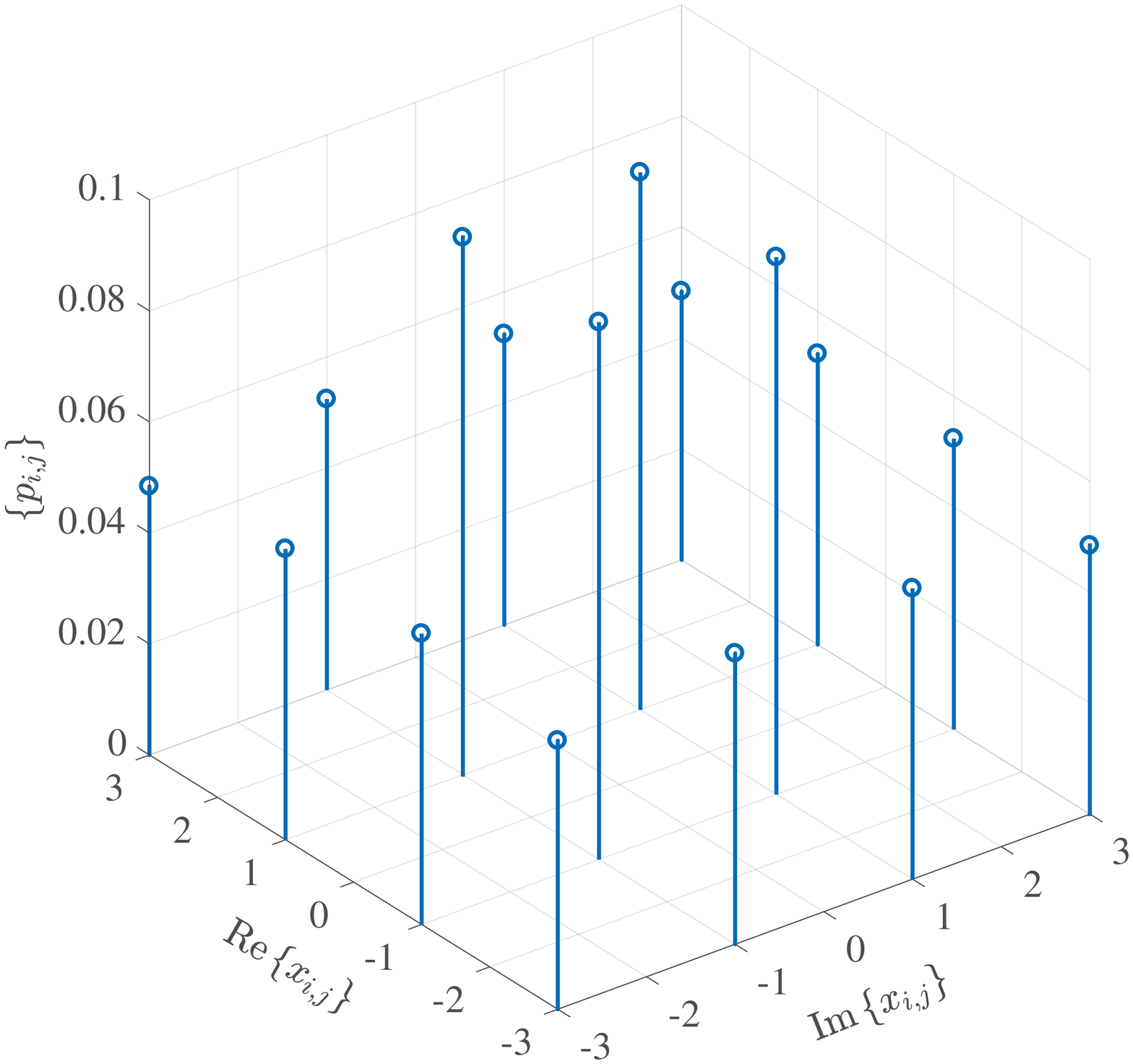}
      \vskip-0.2cm\centering {\footnotesize (a)}
    \end{minipage}
     \begin{minipage}[htbp]{0.45\textwidth}
      \centering
      \includegraphics[height=7.5cm,width=7.5cm]{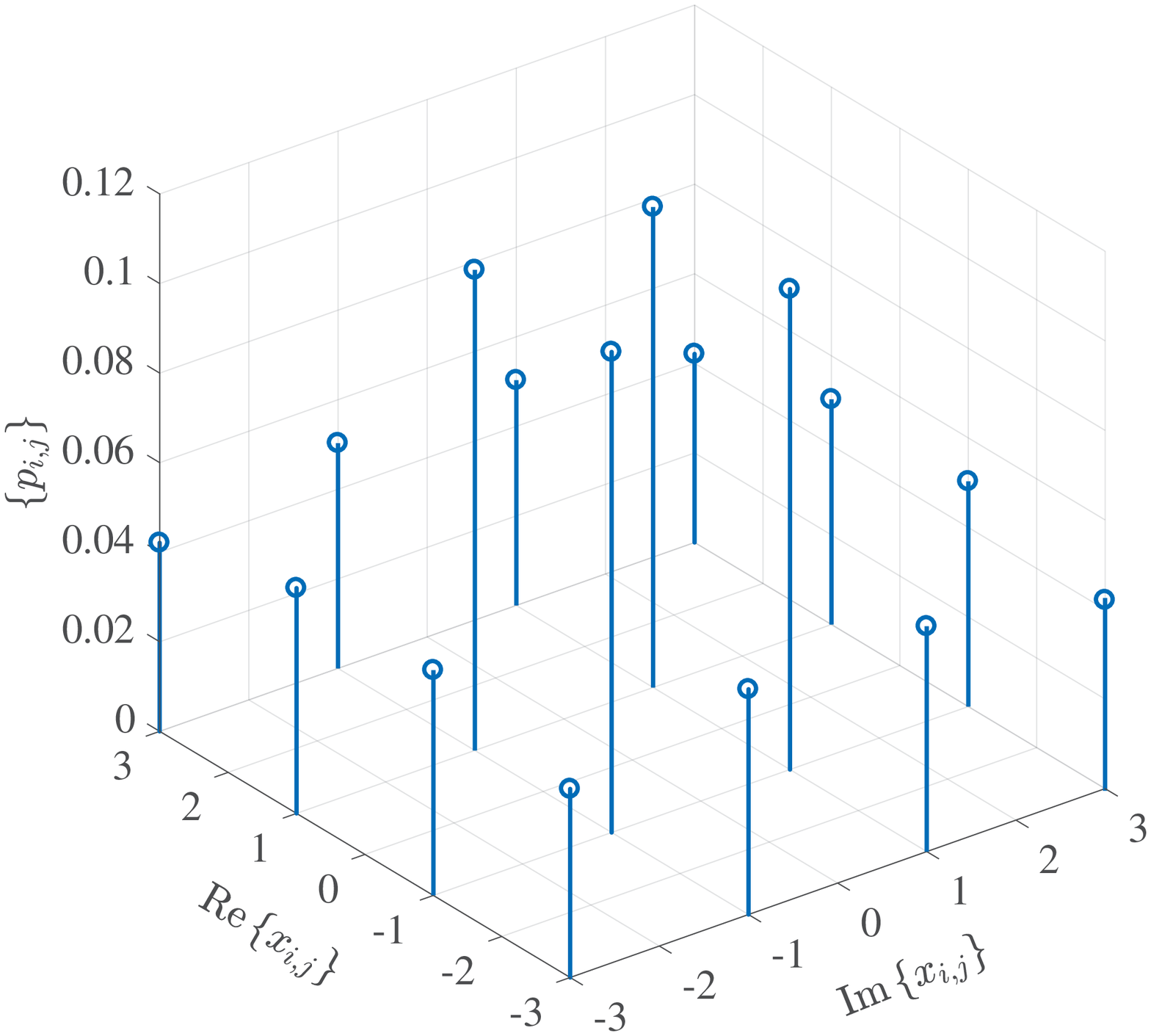}
      \vskip-0.2cm\centering {\footnotesize (b)}
    \end{minipage}\hfill
 \caption{ The  optimal probability of discrete constellation points with  (a) $D\left( {{p_{y,0}}\left\| {{p_{y,1}}} \right.} \right)$  and (b) $D\left( {{p_{y,1}}\left\| {{p_{y,0}}} \right.} \right)$ for the proposed optimal probabilistic constellation shaping design.}
 \label{3}  
\end{figure}
Fig.  \ref{3} shows the optimal probability distribution of
input $\left\{{p_{{i,j}}}\right\}$ with $\rm{SNR=12dB}$ of the proposed optimal  probabilistic constellation shaping design, for $D\left( {{p_{y,0}}\left\| {{p_{y,1}}} \right.} \right)$ in Fig.  \ref{3} (a), and for $D\left( {{p_{y,1}}\left\| {{p_{y,0}}} \right.} \right)$ in Fig.  \ref{3} (b). As it can be seen from  Fig.  \ref{3}, for the proposed optimal  probabilistic constellation shaping design, the optimal probability distribution is not equiprobable, and the symmetrical points have equal probabilities. Specifically, when the number of discrete constellation points is sixteen, the probability of each constellation point for $D\left( {{p_{y,0}}\left\| {{p_{y,1}}} \right.} \right)$ and $D\left( {{p_{y,1}}\left\| {{p_{y,0}}} \right.} \right)$ is given in Table  I. From the table, we can clearly see that when the coordinates of the constellation points are symmetrical, their probabilities are the same, and vice versa.


 \begin{table*}[htbp]
  \centering
  \caption{The optimal probability distribution of the  optimal  probabilistic constellation shaping design for  $D\left( {{p_{y,0}}\left\| {{p_{y,1}}} \right.} \right)$ and $D\left( {{p_{y,1}}\left\| {{p_{y,0}}} \right.} \right)$ with $\rm{SNR=12dB}$ and $K=16$ }
  \begin{tabular}{| m{3.5cm}<{\centering}|m{1cm}<{\centering} | m{1cm}<{\centering} | m{1cm}<{\centering} | m{1cm}<{\centering}|m{1cm}<{\centering} | m{1cm}<{\centering} | m{1cm}<{\centering} | m{1cm}<{\centering}|}
    \hline
     {\multirow{2}*{\diagbox{${\rm{Im}}\left\{{x_{{i,j}}}\right\}$}{$\left\{{p_{{i,j}}}\right\}$}{${\rm{Re}}\left\{{x_{{i,j}}}\right\}$}}}
    &\multicolumn{4}{|c|}{$D\left( {{p_{y,0}}\left\| {{p_{y,1}}} \right.} \right)$} &\multicolumn{4}{|c|}{$D\left( {{p_{y,1}}\left\| {{p_{y,0}}} \right.} \right)$}\\[1.5ex]
    \cline{2-9}
     \multicolumn{1}{|c|}{}&$-3$  &$-1$ &$1$ &$3$ &$-3$  &$-1$ &$1$ &$3$\\[0.8ex]
           \hline
-3&$0.0484$&$0.0524$&$0.0524$&$0.0484$&$0.0454$&$0.0502$&$0.0502$&$0.0454$\\
\hline
-1&$0.0524$&$0.0968$&$0.0968$&$0.0524$&$0.0502$&$0.1041$&$0.1041$&$0.0502$\\
\hline
1&$0.0524$&$0.0968$&$0.0968$&$0.0524$&$0.0502$&$0.1041$&$0.1041$&$0.0502$\\
\hline
3&$0.0484$&$0.0552$&$0.0552$&$0.0484$&$0.0454$&$0.0502$&$0.0502$&$0.0454$\\
\hline
  \end{tabular}
 \end{table*}


\begin{figure}[htpb]
      \centering
	\includegraphics[width=8cm]{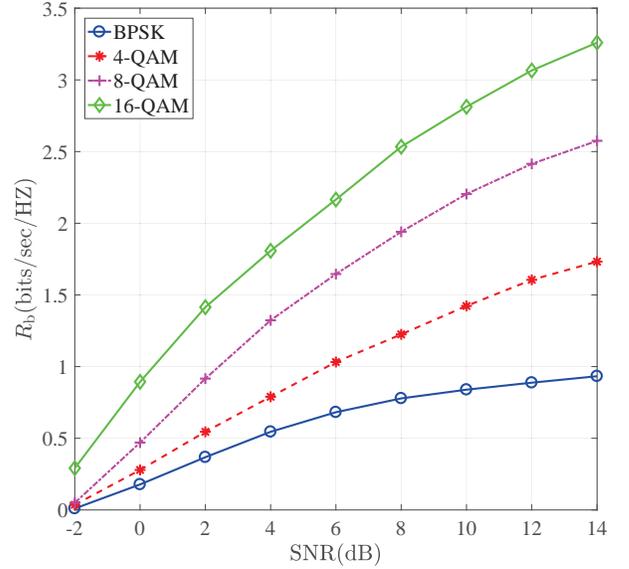}
    \caption{ The achievable covert rate of Bob versus $\rm{SNR}$ with different number of points $K$ for $D\left( {{p_{y,0}}\left\| {{p_{y,1}}} \right.} \right)$.}
  \label{4}
\end{figure}
Fig. \ref{4} considers the proposed optimal  probabilistic constellation shaping design and depicts, the achievable covert rate of Bob ${R_{\rm{b}}}$ versus $\rm{SNR}$ with different number of constellation points
$K = 2, 4, 8, 16$ for $D\left( {{p_{y,0}}\left\| {{p_{y,1}}} \right.} \right)$. It can be seen from
Fig. \ref{4} that when ${\rm{SNR}}$ increases, the covert rate of Bob ${R_{\rm{b}}}$ increases.
{{In addition,  we observe that larger number of points $K$ results in higher covert rate of Bob ${R_{\rm{b}}}$, especially for high SNRs. Thus, as the modulation order increases, the rate of covert communication increases.}}

\begin{figure}[htpb]
      \centering
	\includegraphics[width=8cm]{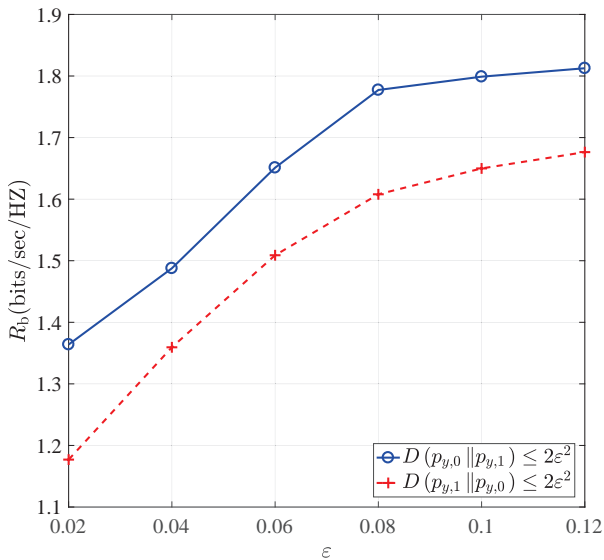}
    \caption{ The covert rate  versus $\varepsilon$  for the proposed optimal  probabilistic constellation shaping design with the cases of $D\left( {{p_{y,0}}\left\| {{p_{y,1}}} \right.} \right)$ and $D\left( {{p_{y,1}}\left\| {{p_{y,0}}} \right.} \right)$.}
  \label{5}
\end{figure}

 Fig. \ref{5} shows the covert rate  versus $\varepsilon$  for the proposed optimal   probabilistic constellation shaping design for $D\left( {{p_{y,0}}\left\| {{p_{y,1}}} \right.} \right)$ and $D\left( {{p_{y,1}}\left\| {{p_{y,0}}} \right.} \right)$, where $K=8$, ${\rm{SNR}}=6{\rm{dB}}$.  It can be observed from the figure that as $\varepsilon$ increases, the covert constraint becomes loose, resulting in a   covert rate increase.  The rate of the covert constraint $D\left( {{p_{y,0}}\left\| {{p_{y,1}}} \right.} \right)\le 2{\varepsilon ^2}$ is higher than that of the case $D\left( {{p_{y,1}}\left\| {{p_{y,0}}} \right.} \right)\le 2{\varepsilon ^2}$.  This is because $D\left( {{p_{y,1}}\left\| {{p_{y,0}}} \right.} \right)$ is less than  $D\left( {{p_{y,1}}\left\| {{p_{y,0}}} \right.} \right)$ for the same probability distribution, and thus $D\left( {{p_{y,1}}\left\| {{p_{y,0}}} \right.} \right)\le 2{\varepsilon ^2}$ is more stringent than $D\left( {{p_{y,0}}\left\| {{p_{y,1}}} \right.} \right)\le 2{\varepsilon ^2}$.

Finally, we compare the performance and complexity of three objective functions.

\begin{figure}
    \begin{minipage}[htbp]{0.45\textwidth}
      \centering
      \includegraphics[height=7.5cm,width=7.5cm]{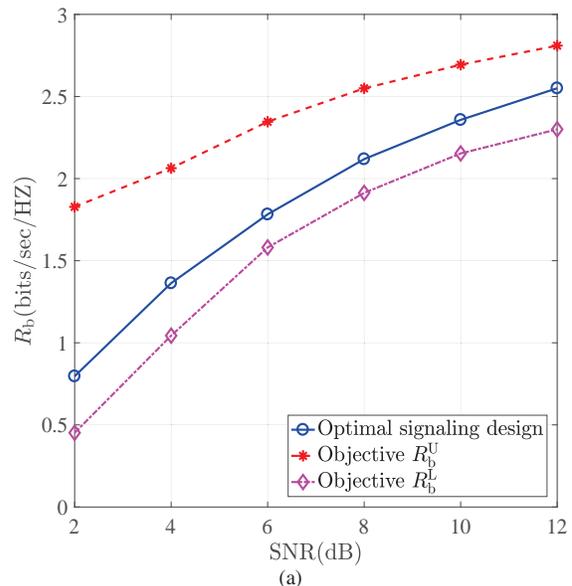}
      \vskip-0.2cm\centering {\footnotesize (a)}
    \end{minipage}
     \begin{minipage}[htbp]{0.45\textwidth}
      \centering
      \includegraphics[height=7.5cm,width=7.5cm]{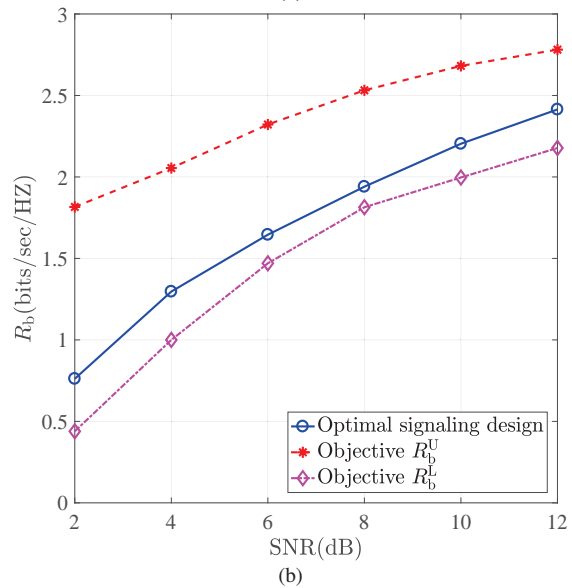}
      \vskip-0.2cm\centering {\footnotesize (b)}
    \end{minipage}\hfill
 \caption{ The  achievable rate of Bob versus $\rm{SNR}$ for  (a) $D\left( {{p_{y,0}}\left\| {{p_{y,1}}} \right.} \right)$  and (b) $D\left( {{p_{y,1}}\left\| {{p_{y,0}}} \right.} \right)$ with the covertness threshold   $2{\varepsilon ^2} = 0.1$.}
 \label{6}  
\end{figure}
Fig. \ref{6} depicts  the achievable rate of Bob ${R_{\rm{b}}}$ with the proposed optimal  probabilistic constellation shaping design, as well as the objective functions ${R_{\rm{b}}^{\text{L}}}$ and ${R_{\rm{b}}^{\text{U}}}$ versus the   ${\rm{SNR}}$  for the case of $D\left( {{p_{y,0}}\left\| {{p_{y,1}}} \right.} \right)$ and $D\left( {{p_{y,1}}\left\| {{p_{y,0}}} \right.} \right)$, respectively.
 It can be observed that the mutual information of Bob ${R_{\rm{b}}}$ increases as
the  ${\rm{SNR}}$ increases, while ${R_{\rm{b}}}$ of the proposed optimal probabilistic constellation shaping design is  between the objective functions ${R_{\rm{b}}^{\text{L}}}$ and ${R_{\rm{b}}^{\text{U}}}$, and the objective function  ${R_{\rm{b}}^{\text{U}}}$ is higher than
 the objective function ${R_{\rm{b}}^{\text{L}}}$.

From   Fig. \ref{6} we observe that the proposed optimal  probabilistic constellation shaping design is  between the objective functions ${R_{\rm{b}}^{\text{L}}}$ and ${R_{\rm{b}}^{\text{U}}}$. Here, we compare the computational complexity of the three designs by computational time
in Table  II, and all simulations of the three methods are performed using MATLAB 2016b with 2.30GHz, 2.29GHz dual CPUs and a 128GB RAM, where $K=8$.
Specifically,  Table  III shows that the computational time of objective functions ${R_{\rm{b}}}$, ${R_{\rm{b}}^{\text{L}}}$ and ${R_{\rm{b}}^{\text{U}}}$ for the covert constraint condition $D\left( {{p_{y,0}}\left\| {{p_{y,1}}} \right.} \right)$ is 55.04, 2.845 and 3.012 seconds, respectively. The computational times of the latter two cases is approximately 95 percent shorter than that of the probabilistic constellation shaping design.
Under the covert constraint condition $D\left( {{p_{y,0}}\left\| {{p_{y,1}}} \right.} \right)$ the computational time of objective functions ${R_{\rm{b}}}$, ${R_{\rm{b}}^{\text{L}}}$ and ${R_{\rm{b}}^{\text{U}}}$ is 107.77, 3.219 and 3.438 seconds, respectively. The computational times of the latter two cases is approximately improved by 97 percent compared to that of the probabilistic constellation shaping design.
  Moreover,
  the computational time of the design for $D\left( {{p_{y,0}}\left\| {{p_{y,1}}} \right.} \right)$ is less than that of $D\left( {{p_{y,1}}\left\| {{p_{y,0}}} \right.} \right)$.
\begin{table}[htbp]
  \centering
  \caption{  Computational time comparison among the objective functions  ${R_{\rm{b}}}$, ${R_{\rm{b}}^{\text{L}}}$ and ${R_{\rm{b}}^{\text{U}}}$}
  \begin{tabular}{|c|c|c|c|c|}
     \hline
    \diagbox{Constraint}{Time/second}{Objective}&${R_{\rm{b}}}$  &${R_{\rm{b}}^{\text{L}}}$ &${R_{\rm{b}}^{\text{U}}}$   \\
   \hline
$D\left( {{p_{y,0}}\left\| {{p_{y,1}}} \right.} \right)$ &$55.04$&$2.845$&$3.012$\\
\hline
$D\left( {{p_{y,1}}\left\| {{p_{y,0}}} \right.} \right)$&$107.77$&$3.219$&$3.438$\\
\hline
  \end{tabular}
 \end{table}

\section{Conclusions}
In this paper, we propose an  optimal  probabilistic constellation shaping design for covert communications,  where Alice
covertly sends a message to Bob while avoiding
being discovered by Willie. We derive the achievable
rate expressions of the covert communications system, and we study the covert rate maximization problem via
optimizing the constellation distribution.
In addition, to strike a balance between the computational complexity and
the transmission performance,
we further develop a framework
that maximizes the upper  and lower bounds  of the achievable rate.
Numerical
results quantify the gains of the proposed beamformers
design over state-of-the-art schemes in terms of the achievable covert rate.

\begin{appendices}
\section{Derivation of the formulation  \eqref{KL_p0p1_ub}}
 The upper bound ${D_{{\rm{U}}}}\left( {{p_{y,0}}\left\| {{p_{y,1}}} \right.} \right)$ on the KL divergence $D\left( {{p_{y,0}}\left\| {{p_{y,1}}} \right.} \right)$ is derived as follows
\begin{subequations}
\begin{align}
&D\left( {{p_{y,0}}\left\| {{p_{y,1}}} \right.} \right)
\le  - \frac{1}{{\ln 2}} \nonumber\\
& - {\log _2}\sum\limits_{k = 1}^K {{p_k}} \exp \left( { - {\mathbb{E}_{{z_{\rm{w}}}}}\left\{ {\frac{{{{\left| {{z_{\rm{w}}} - {g_{\rm{w}}^{\rm{*}}}{x_k}} \right|}^2}}}{{\sigma _{\rm{w}}^2}}} \right\}} \right)\label{Aa}\\
  &=  - \frac{1}{{\ln 2}} - {\log _2}\sum\limits_{k = 1}^K {{p_k}} \exp \left( { - \frac{{{\mathbb{E}_{{z_{\rm{w}}}}}\left\{ {z_{\rm{w}}^2} \right\} + {{\left| {g_{\rm{w}}^{\rm{*}}{x_k}} \right|}^{\rm{2}}}}}{{\sigma _{\rm{w}}^2}}} \right)\\
 & =  - \frac{1}{{\ln 2}} - {\log _2}\sum\limits_{k = 1}^K {{p_k}} \exp \left( { - 1- \frac{{{{\left| {g_{\rm{w}}^{\rm{*}}{x_k}} \right|}^{\rm{2}}}}}{{\sigma _{\rm{w}}^2}}} \right)\\
 & =  - \frac{1}{{\ln 2}} - \left( { - \frac{1}{{\ln 2}} + {{\log }_2}\sum\limits_{k = 1}^K {{p_k}} \exp \left( { - \frac{{{{\left| {g_{\rm{w}}^{\rm{*}}{x_k}} \right|}^{\rm{2}}}}}{{\sigma _{\rm{w}}^2}}} \right)} \right)\\
 & =  - {\log _2}\sum\limits_{k = 1}^K {{p_k}} \exp \left( { - \frac{{{{\left| {g_{\rm{w}}^{\rm{*}}{x_k}} \right|}^{\rm{2}}}}}{{\sigma _{\rm{w}}^2}}} \right),
\end{align}
\end{subequations}
 where inequality \eqref{Aa}  holds due to Jensen's Inequality.

\section{Derivation of the formulation  \eqref{KL_p1p0_ub} }
The upper bound ${D_{{\rm{U}}}}\left( {{p_{y,1}}\left\| {{p_{y,0}}} \right.} \right)$ of the KL divergence $D\left( {{p_{y,1}}\left\| {{p_{y,0}}} \right.} \right)$ is given as
\begin{subequations}
\begin{align}
&D\left( {{p_{y,1}}\left\| {{p_{y,0}}} \right.} \right) \le \sum\limits_{k = 1}^K {{p_k}} {\log _2}\sum\limits_{j = 1}^K {{p_j}}  \nonumber\\
& \times {\mathbb{E}_{{z_{\rm{w}}}}}\left\{ {\exp \left( { - \frac{{{{\left| {g_{\rm{w}}^{\rm{*}}\left( {{x_k} - {x_j}} \right) + {z_{\rm{w}}}} \right|}^2}}}{{\sigma _{\rm{w}}^2}}} \right)} \right\}
+ \frac{1}{{\ln 2}} + \frac{\left|{{ {{g_{\rm{w}}^2}}}{P_{\rm{A}}}}\right|}{{\left( {\ln 2} \right)\sigma _{\rm{w}}^2}}\label{Ba}\\
&  =\sum\limits_{k = 1}^K {{p_k}} {\log _2}\sum\limits_{j = 1}^K {{p_j}} {\mathbb{E}_{{z_{{\rm{w}},R}}}}\left\{ {\exp \left( { - \frac{{{{\left( {{c_R} + {z_{{{\rm{w}},R}}}} \right)}^2}}}{{\sigma _{\rm{w}}^2}}} \right)} \right\}\nonumber\\
 & \times{\mathbb{E}_{{z_{{\rm{w}},I}}}}\left\{ {\exp \left( { - \frac{{{{\left( {{c_I} + {z_{{\rm{w}},I}}} \right)}^2}}}{{\sigma _{\rm{w}}^2}}} \right)} \right\}
+ \frac{1}{{\ln 2}} + \frac{\left|{{ {{g_{\rm{w}}^2}}}{P_{\rm{A}}}}\right|}{{\left( {\ln 2} \right)\sigma _{\rm{w}}^2}}\label{Bb}
  \end{align}
\begin{align}
 &  = \sum\limits_{k = 1}^K {{p_k}} {\log _2}\sum\limits_{j = 1}^K {{p_j}} \left[ {\int\limits_{{\rm{ - }}\infty }^\infty  {\frac{{d{z_{{\rm{w}},R}}}}{{\sqrt \pi  {\sigma _{\rm{w}}}}}\exp \left( { - \frac{{{{\left( {{c_R} + {z_{{\rm{w}},R}}} \right)}^2}}}{{\sigma _{\rm{w}}^2}}} \right.} } \right.\nonumber\\
& \left. {\left. { + \frac{{z_{{\rm{w}},R}^2}}{{\sigma _{\rm{w}}^2}}} \right)} \right] \times \left[ {\int\limits_{{\rm{ - }}\infty }^\infty  {\frac{{d{z_{{\rm{w}},I}}}}{{\sqrt \pi  {\sigma _{\rm{w}}}}}\exp \left( { - \frac{{{{\left( {{c_R} + {z_{{\rm{w}},I}}} \right)}^2} + z_{{\rm{w}},I}^2}}{{\sigma _{\rm{w}}^2}}} \right)} } \right]\nonumber \\
 &+ \frac{1}{{\ln 2}} + \frac{\left|{{ {{g_{\rm{w}}^2}}}{P_{\rm{A}}}}\right|}{{\left( {\ln 2} \right)\sigma _{\rm{w}}^2}}
  \end{align}
\begin{align}
 & = \sum\limits_{k = 1}^K {{p_k}} {\log _2}\sum\limits_{j = 1}^K {{p_j}} \left[ {\frac{{\rm{1}}}{{\sqrt \pi  {\sigma _{\rm{w}}}}}\frac{1}{2}\sqrt {\frac{{\sigma _{\rm{w}}^2\pi }}{2}} \exp \left( { - \frac{{c_R^2}}{{2\sigma _{\rm{w}}^2}}} \right)2} \right]\nonumber\\
& \times \left[ {\frac{{\rm{1}}}{{\sqrt \pi  {\sigma _{\rm{w}}}}}\frac{1}{2}\sqrt {\frac{{\sigma _{\rm{w}}^2\pi }}{2}} \exp \left( { - \frac{{c_I^2}}{{2\sigma _{\rm{w}}^2}}} \right)2} \right]
+ \frac{1}{{\ln 2}} + \frac{\left|{{ {{g_{\rm{w}}^2}}}{P_{\rm{A}}}}\right|}{{\left( {\ln 2} \right)\sigma _{\rm{w}}^2}} \\
 & = \sum\limits_{k = 1}^K {{p_k}} {\log _2}\sum\limits_{j = 1}^K {{p_j}}\frac{1}{2}\exp \left( { - \frac{{c_R^2 + c_I^2}}{{2\sigma _{\rm{w}}^2}}} \right)+ \frac{1}{{\ln 2}} + \frac{\left|{{ {{g_{\rm{w}}^2}}}{P_{\rm{A}}}}\right|}{{\left( {\ln 2} \right)\sigma _{\rm{w}}^2}} \\
  &= \sum\limits_{k = 1}^K {{p_k}} {\log _2}\frac{1}{2}  + \sum\limits_{k = 1}^K {{p_k}} {\log _2}\sum\limits_{j = 1}^K {{p_j}}\exp \left( { - \frac{{{{\left| {g_{\rm{w}}^{\rm{*}}}\left( {{x_k} - {x_j}} \right) \right|}^{\rm{2}}}}}{{2\sigma _{\rm{w}}^2}}} \right)\nonumber\\
& + \frac{1}{{\ln 2}} + \frac{\left|{{ {{g_{\rm{w}}^2}}}{P_{\rm{A}}}}\right|}{{\left( {\ln 2} \right)\sigma _{\rm{w}}^2}} \\
&= \sum\limits_{k = 1}^K {{p_k}} {\log _2}\sum\limits_{j = 1}^K {{p_j}}\exp \left( { - \frac{{{{\left| {g_{\rm{w}}^{\rm{*}}}\left( {{x_k} - {x_j}} \right) \right|}^{\rm{2}}}}}{{2\sigma _{\rm{w}}^2}}} \right)+ \frac{1}{{\ln 2}}\nonumber\\
& + \frac{\left|{{ {{g_{\rm{w}}^2}}}{P_{\rm{A}}}}\right|}{{\left( {\ln 2} \right)\sigma _{\rm{w}}^2}}-1,
\end{align}
\end{subequations}
 where inequality \eqref{Ba}  is true due to  Jensen's Inequality. Equality \eqref{Bb} holds because of the  definitions ${z_{{\rm{w}},R}} \buildrel \Delta \over =  {\rm{Re}}\left\{{{z_{\rm{w}}}} \right\}$ and
${z_{{\rm{w}},I}}\buildrel \Delta \over =  {\rm{Im}}\left\{{{z_{\rm{w}}}} \right\}$,  ${z_{{\rm{w}},R}},{z_{{\rm{w}},I}} \sim {\cal{N}}\left( {0,\frac{1}{2}\sigma _{\rm{w}}^2} \right)$, and  ${c_R}\buildrel \Delta \over ={\mathop{\rm Re}\nolimits} \left( {{g_{\rm{w}}^{\rm{*}}}\left( {{x_k} - {x_j}} \right)} \right)$, ${c_I}\buildrel \Delta \over ={\mathop{\rm Im}\nolimits} \left( {{g_{\rm{w}}^{\rm{*}}}\left( {{x_k} - {x_j}} \right)} \right)$.

   \section{Derivation of the formulation \eqref{mutual_I ub}}
The upper bound ${{R_{\rm{b}}^{{\rm{U}}}}}$ of the covert rate ${R_{\rm{b}}}$ is derived as follows
\begin{subequations}
\begin{align}
&{R_{\rm{b}}}\le  - \sum\limits_{k = 1}^K {{p_k}} {\log _2}\sum\limits_{j = 1}^K {{p_j}}\nonumber\\
& \times \exp \left( { - {\mathbb{E}_{{z_{\rm{b}}}}}\left\{ {\frac{{{{\left| {g_{\rm{b}}^{\rm{*}}\left( {{x_k} - {x_j}} \right) + {z_{\rm{b}}}} \right|}^2}}}{{\sigma _b^2}}} \right\}} \right)
- \frac{1}{{\ln 2}}\label{Ca}\\
 & =   - \sum\limits_{k = 1}^K {{p_k}} {\log _2}\sum\limits_{j = 1}^K {{p_j}} \exp \left( { - \frac{{{{\left| {{g_{\rm{b}}^{\rm{*}}}\left( {{x_k} - {x_j}} \right)} \right|}^2}}}{{\sigma _{\rm{b}}^2}} - \frac{{{\mathbb{E}_{{z_{\rm{b}}}}}\left\{ {\left|z_{\rm{b}}\right|^2} \right\}}}{{\sigma _{\rm{b}}^2}}} \right)\nonumber \\
& - \frac{1}{{\ln 2}}\\
 & =  - \frac{1}{{\ln 2}} - \sum\limits_{k = 1}^K {{p_k}} {\log _2}\sum\limits_{j = 1}^K {{p_j}} \exp \left( { - \frac{{{{\left| {{g_{\rm{b}}^{\rm{*}}}\left( {{x_k} - {x_j}} \right)} \right|}^2}}}{{\sigma _{\rm{b}}^2}} - 1} \right)
   \end{align}
\begin{align}
&  =  - \sum\limits_{k = 1}^K {{p_k}} {\log _2}\sum\limits_{j = 1}^K {{p_j}} \exp \left( { - \frac{{{{\left| {{g_{\rm{b}}^{\rm{*}}}\left( {{x_k} - {x_j}} \right)} \right|}^2}}}{{\sigma _{\rm{b}}^2}}} \right)\exp \left( { - 1} \right)\nonumber \\
&- \frac{1}{{\ln 2}} \\
 & =   - \sum\limits_{k = 1}^K {{p_k}} \left( {{{\log }_2}\sum\limits_{j = 1}^K {{p_j}} \exp \left( { - \frac{{{{\left| {{g_{\rm{b}}^{\rm{*}}}\left( {{x_k} - {x_j}} \right)} \right|}^2}}}{{\sigma _{\rm{b}}^2}}} \right) - \frac{1}{{\ln 2}}} \right)\nonumber\\
& - \frac{1}{{\ln 2}}
   \end{align}
\begin{align}
& =  - \sum\limits_{k = 1}^K {{p_k}} {\log _2}\sum\limits_{j = 1}^K {{p_j}} \exp \left( { - \frac{{{{\left| {{g_{\rm{b}}^{\rm{*}}}\left( {{x_k} - {x_j}} \right)} \right|}^2}}}{{\sigma _{\rm{b}}^2}}} \right),
\end{align}
\end{subequations}
where inequality \eqref{Ca}  is true due to Jensen's Inequality.

  \section{Derivation of the formulation \eqref{mutual_I lb}}
The lower bound ${{R_{\rm{b}}^{{\rm{L}}}}}$ on the covert rate ${R_{\rm{b}}}$ is given as
\begin{subequations}
\begin{align}
&{R_{\rm{b}}} \ge  - \sum\limits_{k = 1}^K {{p_k}} {\log _2}\sum\limits_{j = 1}^K {{p_j}} \nonumber\\
& \times {\mathbb{E}_{{z_{\rm{b}}}}}\left\{ {\exp \left( { - \frac{{{{\left( {g_{\rm{b}}^{\rm{*}}\left( {{x_k} - {x_j}} \right) + {z_{\rm{b}}}} \right)}^2}}}{{\sigma _{\rm{b}}^2}}} \right)} \right\}
- \frac{1}{{\ln 2}} \label{Da}\\
& =  - \sum\limits_{k = 1}^K {{p_k}} {\log _2}\sum\limits_{j = 1}^K {{p_j}} \left[ {\int\limits_{{\rm{ - }}\infty }^\infty  {\frac{{{d{z_{{\rm{b}},R}}}}}{{\sqrt \pi  {\sigma _{\rm{b}}}}}\exp \left( { - \frac{{{{\left( {{a_R} + {z_{{\rm{b}},R}}} \right)}^2} + z_{{\rm{b}},R}^2}}{{\sigma _{\rm{b}}^2}}} \right)} } \right]\nonumber\\
& \times\left[ {\int\limits_{{\rm{ - }}\infty }^\infty  {\frac{{{d{z_{{\rm{b}},I}}}}}{{\sqrt \pi  {\sigma _{\rm{b}}}}}\exp \left( { - \frac{{{{\left( {{a_R} + {z_{{\rm{b}},I}}} \right)}^2} + z_{{\rm{b}},I}^2}}{{\sigma _{\rm{b}}^2}}} \right)} } \right]- \frac{1}{{\ln 2}}\label{Db}\\
&=  - \sum\limits_{k = 1}^K {{p_k}} {\log _2}\sum\limits_{j = 1}^K {{p_j}}\left[ {\frac{{\rm{1}}}{{\sqrt \pi  {\sigma _{\rm{b}}}}}\frac{1}{2}\sqrt {\frac{{\sigma _{\rm{b}}^2\pi }}{2}} \exp \left( { - \frac{{a_R^2}}{{2\sigma _{\rm{b}}^2}}} \right)2} \right]\nonumber\\
& \times\left[ {\frac{{\rm{1}}}{{\sqrt \pi  {\sigma _{\rm{b}}}}}\frac{1}{2}\sqrt {\frac{{\sigma _{\rm{b}}^2\pi }}{2}} \exp \left( { - \frac{{a_I^2}}{{2\sigma _{\rm{b}}^2}}} \right)2} \right] - \frac{1}{{\ln 2}}\\
& =  - \frac{1}{{\ln 2}} - \sum\limits_{k = 1}^K {{p_k}} {\log _2}\sum\limits_{j = 1}^K {{p_j}}\frac{1}{2}\exp \left( { - \frac{{a_R^2 + a_I^2}}{{2\sigma _{\rm{b}}^2}}} \right) \\
 & =- \frac{1}{{\ln 2}}+1 - \sum\limits_{k = 1}^K {{p_k}} {\log _2}\sum\limits_{j = 1}^K {{p_j}}\exp \left( { - \frac{{{{\left| {g_{\rm{b}}^{\rm{*}}}\left( {{x_k} - {x_j}} \right) \right|}^{\rm{2}}}}}{{2\sigma _{\rm{b}}^2}}} \right),
\end{align}
\end{subequations}
where  inequality \eqref{Da}  is true due to  Jensen's inequality, equality \eqref{Db} holds because of the  definitions ${z_{{\rm{b}},R}} \buildrel \Delta \over =  {\rm{Re}}\left\{{{z_{\rm{b}}}} \right\}$,  ${z_{{\rm{b}},I}} \buildrel \Delta \over =  {\rm{Im}}\left\{{{z_{\rm{b}}}} \right\}$,
${z_{{\rm{b}},R}},{z_{{\rm{b}},I}} \sim {\cal{N}}\left( {0,\frac{1}{2}\sigma _{\rm{b}}^2} \right)$, and  ${a_R} \buildrel \Delta \over = {\mathop{\rm Re}\nolimits} \left( {{g_{\rm{b}}^{\rm{*}}}\left( {{x_k} - {x_j}} \right)} \right) $, ${a_I} \buildrel \Delta \over = {\mathop{\rm Im}\nolimits} \left( {{g_{\rm{b}}^{\rm{*}}}\left( {{x_k} - {x_j}} \right)} \right) $.

  \end{appendices}
\bibliographystyle{IEEE-unsorted}
\bibliography{refs0611}

\end{document}